\documentclass[twocolumn,secnumarabic,amssymb, nobibnotes, aps, prd]{revtex4-1}

\usepackage{amsmath}
\usepackage[many]{tcolorbox}
 \usepackage{xcolor}

\usepackage{graphicx}
\usepackage{braket}
\usepackage{natbib}
\usepackage{fancyhdr}
\usepackage{amsthm}

\newtheorem{theorem}{Theorem}
\newtheorem{identity}{Identity}

\begin{document}

\title{Multi-Directional Periodic Driving of a Two-Level System beyond Floquet formalism}%

\author{Michael Warnock}
\email[]{michael.p.warnock3.civ@us.navy.mil}
\affiliation{Department of Physics, Brown University, Providence, Rhode Island 02912, USA}
\affiliation{USW Platforms and Payload Integration Department, Naval Undersea Warfare Center, Newport, RI, 02840 USA}

\author{David A. Hague}%
\affiliation{Sensor and Sonar Systems Department, Naval Undersea Warfare Center, Newport, RI, 02840 USA}

\author{Vesna F. Mitrovi\'c}%
\affiliation{Department of Physics, Brown University, Providence, Rhode Island 02912, USA}

\date{\today}%

\begin{abstract}
\indent In this manuscript, we introduce an exact expression for the response of a semi-classical two-level quantum system subject to arbitrary periodic driving.  Determining the transition probabilities of a two-level system driven by an arbitrary periodic waveform necessitates numerical calculations through methods such as Floquet theory, requiring the truncation of an infinite matrix.  However, such truncation can lead to a loss of significant interference information, hindering quantum sensors or introducing artifacts in quantum control.  To alleviate this issue, we use the $\star$-resolvent formalism with the path-sum theorem to determine the exact series solution to Schr\"odinger's equation, therefore providing the exact transition probability.  The resulting series solution is generated from a compact kernel expression containing all of the information of the periodic drive and then expanded in a non-harmonic Fourier series basis given by the divided difference of complex exponentials with coefficients corresponding to products of generalized Bessel functions.  The present method provides an analytical formulation for quantum sensors and control applications.
\end{abstract}

\maketitle

\thispagestyle{fancy}

\section{Introduction}

\indent  A considerable effort has been spent investigating periodically driven quantum systems \cite{eckardt2017colloquium,rodriguez2021low,goldman2015periodically,sen2021analytic,weitenberg2021tailoring,oka2019floquet,grifoni1998driven,silveri2017quantum,ivakhnenko2023nonadiabatic}.  A few examples include non-trivial topological states \cite{lindner2011floquet, wang2013observation, rudner2013anomalous}, modification of the tunneling between sites \cite{tsuji2011dynamical,gorg2018enhancement}, coherent destruction of tunneling \cite{grifoni1998driven}, dynamical phase transitions \cite{heyl2018dynamical,zvyagin2016dynamical}, discrete time crystals \cite{choi2017observation}, synthetic gauge fields \cite{goldman2014periodically,bukov2015universal}, and higher harmonic generation \cite{gauthey1997high,de2002high}.  
This extensive effort emanates from the potential to transform   quantum control applications with the ultimate goal of developing quantum sensors whose sensitivity is limited only by the fundamental laws of quantum mechanics or realization of quantum gates in quantum computers \cite{yin2021rabi,mishra2021driving,wang2022sensing,wang2021error,shi2016quantum}.  

The simplest quantum sensor can be mathematically represented as    a periodically driven two-level (``qubit'' \cite{nielsen2010quantum}) system. The investigation of the dynamics of the  the periodically driven two-level system is hindered by  the non-commutativity of the corresponding Hamiltonian at differing times,  which impedes the exact solution of the corresponding    Schr\"odinger equation. This non-commutativity leads to set of non-autonomous differential equations whose  solution now depends not only on the elapsed time interval but rather on  the current  and initial times. 
Nevertheless, the problem can be exactly solved in the case of near-resonant single-harmonic drive with circular modulation, as was done by  Rabi to examine  the response of an atom subjected to a single harmonic field with a frequency near the natural frequency of the atom \cite{rabi1937space}.    Extending beyond this initial investigation requires development of numerous methods to address applicability  to general  drive conditions, for example wide regimes of frequency, amplitude, and/or  directions (longitudinal {\it  versus} transverse) of the drive. Most of these methods are applicable in a limited parameter range and/or involve approximations that could introduce unwanted artifacts.   

Here, we introduce an analytical method allowing for the accurate investigation of multi-harmonic, multi-directional periodic driving of a two-level system. Specifically, the general drive is represented by a finite Fourier series in both the transverse and longitudinal directions.  We then introduce the generalized Bessel function (GBF) in order to consolidate the driving to a single term.  This permits us to use the $\star$-resolvent formalism with the path-sum theorem, 
a non-perturbative method for solving systems of non-autonomous differential equations. We obtain explicit equations for the unitary evolution derived from the Schr\"odinger equation that are generated from a compact kernel expression.  The kernel expression generates an exact, uniformly convergent infinite series expanded in a basis given by the divided difference of complex exponentials with products of GBFs as coefficients.  With the exact unitary evolution, the exact transition probability is obtained. 
The expression allows for the accurate investigation of the transition probabilities for quantum sensing and control applications, beyond average Hamiltonian theory \cite{oon2024beyond}.

 For the  periodic driving along the longitudinal direction (diagonal), the exact solutions to the Schr\"odinger equation in terms of Heun's functions  can be obtained for a sole case of a  single harmonic driving field \cite{xie2010analytical,xie2018analytical,ishkhanyan2019quantum}. However, study of the dynamics in the case of the longitudinal drive of more complex form is crucial for providing understanding of Landau-Zener-St\"uckelberg-Majorana (LZSM)  interference phenomena \cite{ivakhnenko2023nonadiabatic, TEDO2025100118}, that 
  arise  when a quantum system is swept across an avoided energy level crossing and forced to undergo  multi-passage at these avoided crossings  \cite{ivakhnenko2023nonadiabatic}. This interference is an essential   tool to efficiently estimate crucial parameters of a quantum system, such as decoherence, coupling strengths, and energy levels. The knowledge of these parameters  enables precise manipulation of a quantum state within a system. 
 When the manipulation requires  large amplitude driving fields, dynamics of the system cannot be solved exactly and approximative schemes are needed.    In particular, in this regime, 
 one can perform a unitary transformation that places the system into a reference frame corresponding to the fast dynamics of the Hamiltonian.  Decomposing the exponentiated periodic drive and invoking the rotating wave approximation (RWA) introduces a time-independent Hamiltonian \cite{fujii2013introduction}.  Similarly, a rate equation approach performs a unitary transformation identical to the RWA, except the transition rate within perturbation theory is introduced.  The resulting dynamics are similar to those from the RWA, except the region of validity is restricted due to the underlying assumption associated with the perturbation theory.  
 There are numerous examples of  investigations in  approximating regimes of driven two-level systems   subject  to complex driving fields. Such examples include  
  certain features such as a single sinusoidal driving \cite{son2009floquet,oliver2005mach,silveri2017quantum,ivakhnenko2023nonadiabatic}, triangular pulses \cite{yan2021lamb}, square wave driving \cite{silveri2015stuckelberg},   biharmonic driving \cite{satanin2014amplitude,blattmann2015qubit,ferron2017mesoscopic},  qubits with bichromatic driving \cite{yan2023multiphoton,forster2015landau}, $N$-step driving fields \cite{shi2021two}, and pulsed periodic driving \cite{deng2016dynamics}.

Alternatively, for periodic driving along the transverse (off-diagonal) direction, analytical solutions   include effective Hamiltonian theories, such as the RWA, and perturbative expansions \cite{liu2022floquet,marinho2024approximate}.  These methods become challenging if one were to increase the tunneling strength beyond the RWA or perturbative regimes \cite{han2024floquet,chen2022enhanced}.  To expand the validity of perturbative regimes, researchers have used the counter-rotating hybridized method \cite{yan2017effects}.  This involves performing a unitary transformation such that the periodic drive is exponentiated, much as it is when considering a longitudinal periodic drive, except that this exponentiated form is split between longitudinal and transverse terms.  Based on the properties of the Bessel function, truncation of the infinite series is performed, allowing for the determination of an effective, time-independent Hamiltonian \cite{han2024floquet}.  

Evidently, the dynamics of a periodically driven two-level system  is sensitive to the waveform shape and the direction of the periodic driving.  However, to determine the dynamics of such a   system  subject to arbitrary driving, one turns to numerical calculations through Floquet theory or Trotterization.  The typical methodology for use of Floquet theory  to address the problem at hand was developed by Shirley \cite{shirley1965solution,ho1983semiclassical,barone1977floquet}. That is, the Hilbert space of the quantum system is extended to a higher dimensional Hilbert space.  The extension of the space allows the reformulation of the time dependent Hamiltonian in terms of a time independent, infinite matrix, known as the Floquet matrix.  Once obtained, diagonalization of a truncated Floquet matrix allows subsequent calculation of the eigenenergies and eigenstates of the quantum system, determining the transition probability.  However, this truncation methodology may lead to various artifacts such as loss of information concerning interference effects or hindering prediction of novel effects generated by the periodic drive.  
 As a result, numerous   novel methods for solving the Schr\"odinger equation with arbitrary drives were explored \cite{rodriguez2021low,sen2021analytic,zeuch2020exact,massa2003new,giscard2023exact,giscard2020dynamics,giscard2015exact,
kalev2021integral,kalev2021quantum,chen2021quantum,neto2023basis}.  A number of these methods are based on  the effective Hamiltonian approach, such as the flow equation  and the exact RWA methodology \cite{zeuch2020exact,massa2003new}.  Others instead seek to determine the unitary evolution by solving the Schr\"odinger equation, such as the method of divided differences applied to the Dyson series, Omega calculus, and the $\star$-resolvent approach \cite{giscard2023exact,giscard2020dynamics,giscard2015exact,
kalev2021integral,kalev2021quantum,chen2021quantum,neto2023basis}.

Therefore, the main goal of this work is to close this gap.  Specifically, we introduce an analytical method allowing for the accurate investigation of multi-harmonic, multi-directional periodic driving of a two-level system.  This manuscript is organized in the following fashion; in section II, we first introduce the two-level system investigated in this manuscript by introducing the multi-tone sinusoidal frequency modulation (MTSFM) to both the transverse and longitudinal directions. The MTSFM has found applications in radar and sonar due to the tunability of its parameters \cite{hague2016generalized,hague2020adaptive,hague2023characterizing}.  In section III, we introduce the $\star$-resolvent formalism with the path-sum theorem for the exact solutions of non-autonomous matrix differential equations.  In section IV, we condense the periodic driving to a single term given by the generalized Bessel functions (GBF) and determine an exact series expression in terms of GBFs and exponential divided differences, providing the exact transition probability.  In section V, we discuss truncation of our exact results.  In section VI, we provide several applications and future areas of investigation to which this work can contribute.  Finally, in section VII, we conclude this manuscript.

\section{Semiclassical Rabi Model}

We present a variant of the Rabi model describing the response of a two-level atom subjected to an external electromagnetic field.  In a semi-classical regime, the atom is quantized to allow two energy levels while the electromagnetic field interaction with the atom is a continuous field corresponding to the energy flow between the atom and the field, as has been previously investigated \cite{son2009floquet,oliver2005mach,silveri2017quantum,satanin2014amplitude,ivakhnenko2023nonadiabatic}.  The electromagnetic field generates an energy bias and a tunnel level splitting given by modulation in both the longitudinal and transverse directions, respectively.  Explicitly, the Hamiltonian is given by the following

\begin{equation}
H(t)= \frac{1}{2} \begin{pmatrix}
\epsilon(t) & \Delta(t) \\ 
\overline{\Delta}(t) & -\epsilon(t)
\end{pmatrix} \text{,}
\end{equation}

\noindent where $\epsilon(t) = \epsilon_0 + \epsilon_{ac}(t) $ is the longitudinal driving modulating the energy bias of the qubit and $\Delta(t)=\Delta + \Delta_{ac}(t)$ is the transverse driving modulating the tunnel level splitting, and $\overline{\bullet}$ denotes complex conjugation.  In the static case, the potential energy of the qubit becomes a double well potential with quantum mechanical tunneling causing the appearance of two discrete levels.  The eigenvalues for this static case are given by $E_0 = -\sqrt{\epsilon_0^2 + \Delta^2}/2$ and $E_1 = \sqrt{\epsilon_0^2 + \Delta^2}/2$ , with the corresponding basis vectors $\ket{0} = 1/\sqrt{2} \left(1, -1 \right)^{T} $  and $\ket{1} = 1/\sqrt{2} \left(1, 1 \right)^{T} $, respectively.  The modulating components of the time-dependent terms are

\begin{subequations}
\begin{equation}
\epsilon_{ac}(t) = \sum_{n=1}^N A_n \cos(n\omega_\epsilon t) + \sum_{m=1}^M B_m \sin(m\omega_\epsilon t)\textit{;}
\end{equation}
\begin{equation}
\Delta_{ac}(t) = \sum_{k=-K}^K \Delta_k e^{i k \omega_\Delta t}\textit{,}
\end{equation}
\end{subequations}

\noindent where $A_n$ ($B_m$) is the Fourier coefficient of the $n$-th ($m-$th) even (odd) harmonic of frequency $\omega_\epsilon$; $N$ is the number of driving frequencies in the longitudinal direction for the even harmonic; $M$ is the number of driving frequencies in the longitudinal direction for the odd harmonic; $\Delta_k$ are the Fourier coefficients of the transverse modulation of frequency $\omega_\Delta$; and, $K$ is the number of harmonics in the transverse direction.  We note the majority of the literature assumes a modulation given by cosines with differing phases to break time reversal symmetry, however, in order to maintain contact with the theoretically investigated and experimentally validated multi-tone sinusoidal frequency modulation (MTSFM) in sonar and radar applications we forego this notation and continue with the modulations as shown in Eq. (2).  Lastly, this is an approximation of a more general quantum system.  While we maintain the semi-classical regime for this investigation, more general investigations have been carried out with quantum interactions between the field and the atom \cite{ivakhnenko2023nonadiabatic,aravind1984two,chu1989generalized}.  The primary difference between this investigation with those performed previously is that, in previous investigations, the electromagnetic field is quantized.  This allows one to consider quantum statistical differences as individual photons interact with the two-level system whereas we assume the number of photons is indistinguishable from a classical field, therefore we are not required to consider individual photon statistics.  Maintaining the considerations above and the approximation of a classical field allows us to use common methods for solving coupled differential equations to determine the response of our system.  In particular, we use the $\star$-resolvent formalism to determine the exact evolution in subsequent sections.  The $\star$-resolvent formalism reduces the coupled system into a scalar function expanded in $\star$-powers using a $\star$-product defined below.  We note that this method is not related to the Moyal $\star$-product where the Moyal $\star$-product is commonly used for investigating quantum systems in phase space with non-commuting operators.  The $\star$-product as it is used here is an integral equation commonly found in the literature associated with the exact solutions of differential equations and Volterra integral equations \cite{giscard2023exact,giscard2020dynamics,giscard2015exact, brunner2017volterra}.

\section{Algebraic Walk Theory}

We use the $\star$-resolvent formalism by exploiting the path-sum theorem as presented in \cite{giscard2023exact,giscard2020dynamics,giscard2015exact}.  While there have been numerous methods of determining exact analytical results for specific waveforms, expanding these results beyond specifics is a challenge.  By exploiting the path-sum theorem, P.L. Giscard, \textit{et. al.}, were able to determine the exact evolution of the time-ordered exponential \cite{giscard2015exact}.  The method exploits the path-sum theorem involving the counting of all the simple cycles and paths on the adjacency graph corresponding to the time-dependent Hamiltonian.  The primary motivation of determining an analytical result is by the observation that several well-established special functions are the solutions to differential equations, such as Bessel's or Heun's differential equations.  However, the solutions to these differential equations are none other than infinite series given a specific name due to their prevalence in various scientific communities.  Based on this observation, algebraic walk theory determines an infinite series solution to the time-ordered exponential that uniformly converges to the exact solution at a super-exponential rate within some defined time interval.  

Within the context of exact solutions to non-autonomous differential equations, the $\star$-resolvent approach revolves around the exact solutions of non-autonomous differential equations.  The solutions to the differential equations are expressed in terms of a generalized $\star$-product represented by a non-commutative, convolution like integral

\begin{equation}
(f \star g)(t,s) = \int_s^t f(t,\tau)g(\tau,s)\, d\tau \textit{.}
\end{equation}

Then, a matrix differential equation may be rewritten as a two-variable system

\begin{subequations}
\begin{equation}
\frac{d}{dt} U(t,s)=A(t)U(t,s)
\end{equation}
\begin{equation}
\frac{d}{dt} U(t,s) \Theta(t-s)=A(t,s)U(t,s)\Theta(t-s) \textit{.}
\end{equation}
\end{subequations}

Defining the Green's function as $\frac{d}{dt} U(t)\Theta(t-s)+Id_\star$, and by properties of the $\star$-product, the system of differential equations may be written as $A(t,s)U(t,s)\Theta(t-s)=A \star G$, where $\Theta(t-s)$ denotes the Heaviside Theta function.  Based on this observation, the Green's function is 

\begin{equation}
G(t,s) = (1_\star - A(t,s)\Theta(t-s))^{\star -1} \textit{.}
\end{equation}

\noindent Here $1_\star$ denotes the Dirac Delta distribution.  Then, the Green's function is the $\star$-resolvent of the two-time matrix $A(t,s)$ and may be expanded in terms of an unconditionally convergent Neumann series.  Finally, the solution to the original system of differential equations results in $U=\Theta \star G$, where one may now invoke the path-sum theorem allowing for the resummation of infinitely many terms as a $\star$-resolvent of scalar functions.
 
In the case of a driven two-level system presented here, the time-ordered exponential is given by the solution to the differential equation 

\begin{equation}
\frac{d}{dt} U(t,s) = -i H(t) U(t,s) \textit{,}
\end{equation}

\noindent with an arbitrary $2 \times 2$ Hamiltonian.  The dynamical adjacency graph corresponding to the two-level system is a fully connected $K_2$ graph, with self-loops.  Evaluating the time-ordered exponential using the path-sum theorem, the exact components of the unitary matrix are

\begin{subequations}
\begin{equation}
U_{11} (t,s) = \int_s^t G_{11}(\tau,s) \, d\tau \textit{;}
\end{equation}
\begin{equation}
U_{22} (t,s) = \int_s^t G_{22}(\tau,s) \, d\tau \textit{;}
\end{equation}
\begin{equation}
U_{21} (t,s) = \int_s^t G_{11/\left\lbrace2\right\rbrace}(t,\tau)\star \overline{\Delta}(\tau) \star G_{22}(\tau,s) \, d\tau \textit{;}
\end{equation}
\begin{equation}
U_{12} (t,s) = \int_s^t G_{22/\left\lbrace1\right\rbrace}(t,\tau)\star \Delta(\tau) \star G_{11}(\tau,s) \, d\tau \textit{.}
\end{equation}
\end{subequations}

\noindent where the Green's functions are given by 

\begin{subequations}
\begin{equation}
G_{11}(t,s) = (1_\star - i\epsilon - \overline{\Delta} \star G_{22/\left\lbrace1\right\rbrace} \star \Delta)^{\star-1}\text{;}
\end{equation}
\begin{equation}
G_{22}(t,s) = (1_\star + i\epsilon + \Delta \star G_{11/\left\lbrace2\right\rbrace} \star \overline{\Delta})^{\star-1}\text{;}
\end{equation}
\begin{equation}
G_{11/\left\lbrace2\right\rbrace}(t,s) = (1_\star - i\epsilon)^{\star-1}(t,s)\text{;}
\end{equation}
\begin{equation}
G_{22/\left\lbrace1\right\rbrace}(t,s) = (1_\star + i\epsilon)^{\star-1}(t,s)\text{.}
\end{equation}
\end{subequations}

\noindent Here the energy biases and tunneling terms are time-dependent.  The arguments of the Green's function are the kernels representing how the system behaves based on changes in the two time variables.  The infinite series solution is acquired by Neumann expansion expanded as a series of  $\star$-products

\begin{subequations}
\begin{equation}
G_{ii}(t,s) = \delta(t-s) + \sum_{k=1}^\infty K_{ii}^{\star k} (t,s) \text{;}
\end{equation}
\begin{equation}
K_{ii}^{\star k} (t,s) = \int_s^t \dots \int_{\tau_{k-1}}^t K_{ii}(t,\tau_{k}) \dots K_{ii}(\tau_1,s) \, d\tau_{k} \dots \, d\tau_1 \text{.}
\end{equation}
\end{subequations}

\noindent Once the Neumann series is evaluated and the unitary is acquired, the exact transition probability is $p(t,s)=|U_{12}(t,s)|^2$.

\section{Exact Transition Probability}

To determine the transition probability, we perform a similar unitary transformation to previous work \cite{silveri2017quantum,ivakhnenko2023nonadiabatic,son2009floquet,oliver2005mach,yan2021lamb,silveri2015stuckelberg,satanin2014amplitude}.  We transform the Hamiltonian to a rotating reference frame using the unitary
	
\begin{equation}
\ket{\Psi(t)}=U_0(t) \ket{\chi(t)}\textit{,}
\end{equation}

\noindent where
	
\begin{equation}
U_0(t) = e^{-\frac{i}{2}(\epsilon_0 t + \sum_{n=1}^N \frac{A_n}{n \omega_\epsilon} \cos(n\omega_\epsilon t) + \sum_{m=1}^M \frac{B_m}{m \omega_\epsilon} \sin(m\omega_\epsilon t))\sigma_z}
\end{equation}

\noindent with $\sigma_z$ being the Pauli matrix.  The Schr\"odinger equation after transformation takes the form ($\hbar = 1$)

\begin{multline}
i \frac{\partial}{\partial t} \ket{\chi (t)} \\ = \left(U_0^\dag H(t) U_0 - i U_0^\dag \frac{\partial U_0 }{\partial t}\right)\ket{\chi(t)} =  H_{rot}(t) \ket{\chi(t)} \textit{.}
\end{multline}

After calculation the modified Hamiltonian, $H_{rot}$, is explicitly written

\begin{subequations}
\begin{equation}
H_{rot}(t)= \\
\frac{1}{2} \begin{pmatrix}
0 & \Delta_{rot}(t)\\ 
\overline{\Delta}_{rot}(t) & 0
\end{pmatrix} \text{;}
\end{equation}
\begin{equation}
\Delta_{rot}(t)= \sum_{k=-K}^K \Delta_k e^{i k \omega_\Delta t} \sum_{l=-\infty}^\infty \mathcal{J}_{l}^{\substack{1:N \\ 1:M}} e^{i l \omega_\epsilon t} \text{,}
\end{equation}
\end{subequations}

\noindent where a generalized version of the Jacobi-Anger relation was used to introduce the generalized Bessel functions (GBF) \cite{dattoli1996theory} with arguments suppressed in the above but given by $\left\lbrace A_n/\omega \right\rbrace= \left\lbrace A_1/\omega,A_2/2\omega,A_3/3\omega,\dots,A_n/n\omega \right\rbrace$ and $\left\lbrace(iB_m)/\omega\right\rbrace=\left\lbrace(iB_1)/\omega,(iB_2)/2\omega,(iB_3)/3\omega,\dots,(iB_m)/m\omega\right\rbrace$ representing vectors of modulation indices consolidating the information of the longitudinal modulation to a single term. Assuming commensurate frequencies for the modulations, the primary frequency is found by $\omega = \text{gcd}(\omega_\epsilon, \omega_\Delta)$.  Then, we obtain the rotated Hamiltonian

\begin{subequations}
\begin{equation}
H(t)= \begin{pmatrix}
0 & \sum_{l=-\infty}^\infty \mathcal{J}_l^\Delta e^{i (\epsilon_0 + l \omega) t} \\ 
\sum_{l=-\infty}^\infty \overline{\mathcal{J}_l^\Delta} e^{-i (\epsilon_0 + l \omega) t} & 0
\end{pmatrix} \text{;}
\end{equation}
\begin{equation}
\mathcal{J}_l^\Delta = \sum_{k=-K}^K \frac{\Delta_k}{2} \mathcal{J}_{l-k}^{\substack{1:N \\ 1:M}}\left(\left\lbrace \frac{A_n}{\omega}\right\rbrace,\left\lbrace \frac{i B_n}{\omega} \right\rbrace \right) \text{.}
\end{equation}
\end{subequations}

The final Hamiltonian consists of a complex Fourier series with amplitudes given by weighted GBFs.  Truncating the infinite series would provide an approximate solution, however, proper choice of truncation can be made by using Carson's bandwidth rule \cite{carson2006notes} and properties of the GBFs \cite{kuklinski2019identities}.  In particular, one may truncate the infinite summation based upon the waveform and energy concentrated in the waveform, or one may perform common perturbative methods.  Instead, we seek the exact transition probability corresponding to Eq. (14).  

Substituting the corresponding Hamiltonian elements into Eq. (8) the Green's functions are reduced to

\begin{subequations}
\begin{equation}
G_{11}(t,s) = (1_\star - \overline{\Delta} \star \Delta)^{\star-1}(t,s) \text{;}
\end{equation}
\begin{equation}
G_{22}(t,s) = (1_\star + \Delta \star \overline{\Delta})^{\star-1}(t,s) \text{.}
\end{equation}
\end{subequations}

\noindent Explicitly evaluating the $\star$-product, $\overline{\Delta} \star \Delta$, results in the expression for the kernel

\begin{equation}
K(t,s) = \sum_{m,n} \overline{\mathcal{J}_n^\Delta} \mathcal{J}_m^\Delta e^{i f_m t - i \frac{f_n}{2} (t + s)} (t-s) \text{sinc}\left(\frac{f_n}{2}(t-s)\right) \textit{,}
\end{equation}

\noindent where $f_l=\epsilon_0+l \omega$.  The $\star$-powers of the Neumann series in Eq. (9b) are multi-dimensional integrals over a simplex \cite{wolpert1995estimating,lasserre2001solving,robins2021friendly}.   Integrals of this form have been carried out using the Hermite-Genocchi formula \cite{kalev2021integral,kalev2021quantum,chen2021quantum} or Omega calculus \cite{neto2023basis}, finding the solution in terms of the exponential divided difference.  Alternatively, similar integrals are related to generalized ambiguity functions resulting in a recursive series solution involving sinc and complex exponential \cite{lu2009computable}.  Or, they may be evaluated as the Fourier transform of the Dirichlet distribution with specific parameters, resulting in multi-dimensional confluent hypergeometric functions \cite{dello2019characteristic}. 

We continue with the divided difference method due to the compactness of the resulting expressions and the ease with which the properties of the divided difference may be exploited.  The kernel given in Eq. (16) may be re-written in terms of the first exponential divided difference as 

\begin{equation}
K(t,s) = (-i) \sum_{m,n} \overline{\mathcal{J}_n^\Delta} \mathcal{J}_m^\Delta e^{i (m\omega - n \omega) t} e^{i[f_n,0](t-s)} \textit{,}
\end{equation}

\noindent where the notation here for divided difference corresponds to the following

\begin{equation}
e^{i[\omega_n, \omega_{n-1},...,\omega_0]t} = \sum_{k=0}^n \frac{e^{i\omega_k t}}{\prod_{k \neq j} (\omega_k - \omega_j)} \textit{.}
\end{equation}

\noindent If a subset of the arguments in the above expression are equivalent, a limiting procedure is required and can be analytically written as

\begin{multline}
f[\overbrace{x_0,x_0,\dots,x_0}^{k_0 + 1},\dots,\overbrace{x_n,x_n,\dots,x_n}^{k_n + 1}] = \\ \frac{1}{k_0!...k_n!} \frac{\partial^{k_0}}{\partial x_0^{k_0}} \dots \frac{\partial^{k_n}}{\partial x_n^{k_n}} f[x_0,\dots,x_n] \textit{.}
\end{multline}

\noindent Continuing with Eq. (17) the $k$-th term of the Neumann series is of the form

\begin{multline}
K^{\star k} (t,s) = \sum_{\left\lbrace m,n \right\rbrace} \overline{\mathcal{J}_{\left\lbrace n \right\rbrace}^\Delta} \mathcal{J}_{\left\lbrace m \right\rbrace}^\Delta \int_s^t \dots \int_{\tau_{k-1}}^t \\ \times e^{i(m_1 \omega - n_1 \omega) t} e^{i[f_{n_{1}},0](t-\tau_k)} \dots \\
 \times e^{i(m_k \omega - n_k \omega)\tau_1} e^{i[f_{n_{k}},0](\tau_1-s)} \, d\tau_{k} \dots \, d\tau_1 \textit{.}
\end{multline}

\noindent Here the summation is over multiple indices of $m_i$ and $n_i$ for $i \in \mathbb{Z}_+$.  As shown in Appendix A, after using a variant of the Hermite-Gnocchi formula \cite{baxterexponential}, Eq. (20) may be rewritten as

\begin{multline}
K^{\star k} (t,s) = (-i)^{2k+1} \sum_{\left\lbrace m,n \right\rbrace} \overline{\mathcal{J}_{\left\lbrace n \right\rbrace}^\Delta} \mathcal{J}_{\left\lbrace m \right\rbrace}^\Delta e^{i(M_k-N_k)t} \\ \times e^{i[\epsilon_0 - M_{k-1} + N_k, -M_{k-2} + N_{k-1}, \dots, \epsilon_0 + M_1, 0](t-s)} \textit{,}
\end{multline}

\noindent where $M_j = \Sigma_{i=1}^{j} m_i \omega$, $N_j = \Sigma_{i=1}^{j} n_i \omega$, and there are $2k$ terms in the argument of the exponential divided difference.  With access to the $k$-th term, the components of the unitary matrix are obtained from Eq. (7).  

\begin{subequations}
\begin{multline}
U_{11}(t,s) = 
1 + \sum_{\left\lbrace m,n \right\rbrace} \overline{\mathcal{J}_{\left\lbrace n \right\rbrace}^\Delta} \mathcal{J}_{\left\lbrace m \right\rbrace}^\Delta e^{i(M_k-N_k)s} \\ \times e^{i[\epsilon_0 - M_{k-1} + N_k, -M_{k-2} + N_{k-1}, \dots, \epsilon_0 + M_1, 0](t-s)} \text{;}
\end{multline}
\begin{multline}
U_{12}(t,s) = 
 \sum_{\left\lbrace m,n \right\rbrace} \overline{\mathcal{J}_{\left\lbrace n \right\rbrace}^\Delta} \mathcal{J}_{\left\lbrace m+1 \right\rbrace}^\Delta e^{i(\epsilon_0+M_k-N_{k+1})s} \\ \times e^{i[\epsilon_0+M_k-N_{k+1}, -M_{k-2} + N_{k-1}, \dots, \epsilon_0 + N_1, 0](t-s)} \text{,}
\end{multline}
\end{subequations}

\noindent with $U_{22} (t,s) = -U_{11}(t,s)$ and $U_{21} (t,s) = -\overline{U_{12}}(t,s)$, providing our main result.  By construction of the approach used, the above expressions provide the exact unitary evolution for a two-level system with arbitrary periodic driving in some time interval, therefore providing the exact transition probability.  The resulting expressions are similar to those found in the literature for Hamiltonian simulation \cite{kalev2021integral,kalev2021quantum} and Omega calculus methods \cite{neto2023basis}, however, in this investigation, the resulting series expressions are derived from a single two-time kernel that compactly contains all of the information of the general periodic driving.  This expression is exact for all parameter regimes and driving conditions in a semi-classical periodically driven two-level system.  

\begin{figure*}[t]
    \centering
    \includegraphics[width=0.8\textwidth]{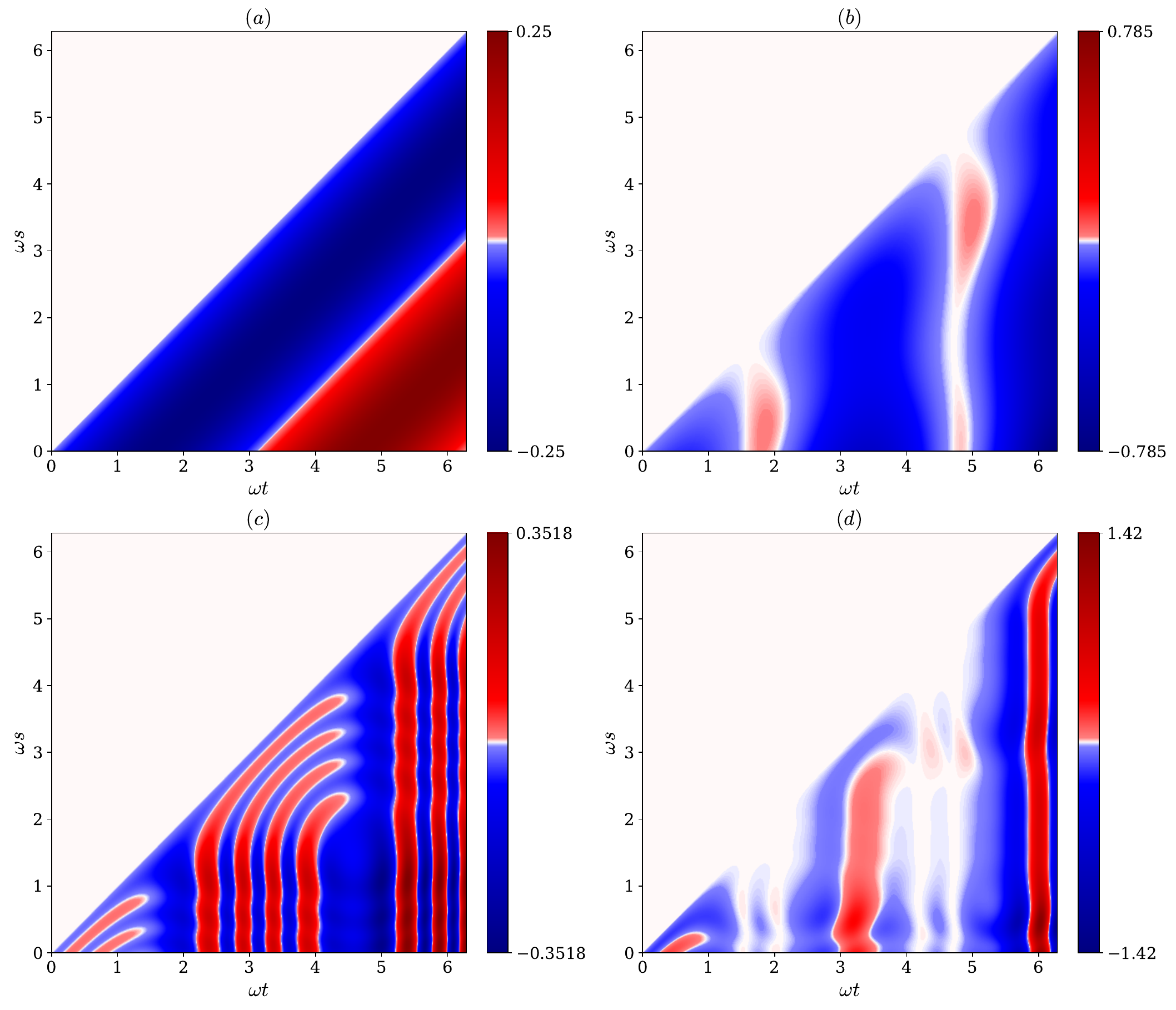}
    \caption{\textbf{Two-Time Kernel}:  Plots of the real part of the kernel in the two-time domain given by $[0,T] \cup [0,T]$ for period $T=2\pi/\omega$ where the abscissa is the first time variable and the ordinate is the second time variable.  The waveforms used are (a) no periodic driving, (b) periodic driving corresponding to the usual Bloch-Siegert Hamiltonian with a tunneling strength of $\Delta = 0.5\omega$, (c) periodic driving corresponding to the usual single harmonically driven two-level system in the longitudinal direction with a tunneling strength of $\Delta = 0.5\omega$ and amplitude of $A = 15\omega$, and (d) periodic driving corresponding to a longitudinal modulation with the first and second harmonics with amplitudes $A_1 = 10\omega$ and $A_2 = 20\omega$, respectively, a constant energy bias $\epsilon_0 = 10\omega$, and transverse modulations given by the first and second harmonics with amplitudes  $\Delta_i = 0.25\omega$.  As expected, the kernel with no periodic driving suggests that time invariance is not broken, as expected with the information contained in the kernel reducible to a single dimension.  However, as soon as periodic driving is introduced, time-invariance is broken and the information contained in the kernel is fully two-dimensional displaying a feature of non-autonomous systems.}
    \label{fig:wide}
\end{figure*}

\subsection{Examination of the Kernel}

It is clear that the dynamics of the system are dependent upon the kernel in Eqs. (16) or (17) and its $\star$-powers generating the series expansion that the unitary is derived from.  Motivated by this, we plot the kernel for several driving conditions in Figure 1, where the triagonal domain denotes the causality of the kernel that is enforced by the Heaviside Theta function.  Fig. 1(a) denotes a kernel with no periodic driving, Fig. 1(b) displays the kernel of the usual Bloch-Siegert Hamiltonian, Fig. 1(c) displays a kernel for a Hamiltonian with the first harmonic in the longitudinal direction representing the usual frequency modulated two-level system, and Fig. 1(d) displays a kernel for the first and second harmonic in both the longitudinal and transverse directions.  Observation of the various kernels shows when time-translation invariance is broken.  When there is no periodic driving, the kernel in Fig. 1(a) is symmetric in the two-time variables, however, invariance is broken with any driving, as one would expect due to the non-commutativity of the corresponding Hamiltonians at differing times.  A commutative Hamiltonian would find that the $\star$-product corresponds to a convolutional product, suggesting that the functions are invariant under time translations.  However, the non-commutativity and lack of time invariance in the structure of the kernel displays the need for the general, non-convolutional product provided in Eq. (3).

Additional observations of the kernel can be determined upon examination of Eqs. (16) or (17).  We separate the kernel based on the magnitude of the energy bias term by letting $\epsilon_0 = -\alpha \omega$ for some integer $\alpha$, which we call integer resonance.  This corresponds to a resonance at which the sinc function is equal to one, resulting in the kernel

\begin{multline}
\overbrace{\overline{\mathcal{J}_\alpha^\Delta} \mathcal{J}_\alpha^\Delta (t-s)}^{\text{RWA}} + \overbrace{\sum_{\substack{m \\ m \neq \alpha}} \overline{\mathcal{J}_{\alpha}^\Delta} \mathcal{J}_m^\Delta e^{i(m\omega-\alpha\omega)t} (t-s)}^{\text{CR}} + \\ \overbrace{\sum_{\substack{m, n \\ m \neq \alpha \\ n \neq \alpha}} \overline{\mathcal{J}_n^\Delta} \mathcal{J}_m^\Delta e^{i f_m t - i \frac{f_n}{2} (t + s)} \text{sinc}\left(\frac{f_n}{2}(t-s)\right)}^{\text{OR}} \text{,}
\end{multline}

\noindent where we denote the terms by the RWA, counter-rotating (CR), and off-resonant (OR) terms.  Ignoring the counter-rotating and the off-resonant terms, the RWA term would remain.  The resulting unitary would correspond to $\star$-powers of this term alone and result in a series expansion corresponding to the Taylor series of the cosine and sine functions.  However, returning to Eq. (23), it isn't  clear at what parameter values one is able to disregard the counter-rotating and off-resonant terms.  Averaging the kernel over a single period in a triangular domain results in 

\begin{multline}
\frac{2}{T^2}\int_0^T \int_0^t K(t,s) \, ds \, dt = \\ \frac{2}{T^2} \sum_{m,n} \overline{\mathcal{J}_n^\Delta} \mathcal{J}_m^\Delta e^{i[\epsilon_0 + m\omega, m\omega -n \omega, m\omega - n\omega, 0]T} \text{.}
\end{multline}

\noindent Here we can separate out the divided difference term based on the indices of the summations.  The resonant case corresponds to the arguments of the exponential divided difference all equal to zero representing the maximum contribution to the average of the kernel with interference patterns that correspond to the GBFs.  The counter-rotating terms can similarly be examined when $\epsilon_0 \neq -\alpha \omega$ where there is constructive interference of the driving at $m=n$ resulting in an equality of three of the arguments of the divided difference.  If $\epsilon_0 = -\alpha \omega$, and using Eq. (24), the CR terms are

\begin{multline}
\frac{2}{T^2} \sum_{\substack{m \\ m=n \\ m \neq \alpha}} \overline{\mathcal{J}_m^\Delta} \mathcal{J}_m^\Delta e^{i[\epsilon_0 + m\omega, 0, 0, 0]T} = \\ \sum_{\substack{m \\ m=n \\ m \neq \alpha}} \overline{\mathcal{J}_m^\Delta} \mathcal{J}_m^\Delta \left(\frac{4i}{(m\omega - \alpha\omega)^2 T} - \frac{2}{m \omega - \alpha \omega} \right) \text{.}
\end{multline}

\noindent The contributions are inversely proportional to the frequency, with a magnitude that is dependent on the separation of the energy bias to integer multiples of the frequency, which is one understanding of the emergent dynamics of the RWA as the average integration of a complex exponential, $e^{in \theta}$, for some $n \in \mathbb{Z}$ decreases like $1/n$ resulting in higher frequency contributions becoming negligible \cite{fujii2013introduction}.  Inverse frequency expansions are common for the investigations of periodically driven systems therefore it is natural that an inverse frequency relationship appears for the contributions to the kernel.  If the energy bias is not some integer multiple of the frequency, or off integer resonance, then the resulting kernel has contributions from sinc functions that are inversely proportional to the difference between the energy bias and some integer multiple.  The dominant contributions would come from the modes in which the energy bias is closest, resulting in a mixture of interference patterns of the GBFs in the averaged kernel and the dynamics.  In the context of quantum sensing, the kernel shows that periodic driving acts as a filter.  The frequencies that dominate the probability for long times are those that will correspond to a resonance, or the RWA term in Eq. (23).  The CR terms can have a non-negligible impact on the probability and may be used to induce additional oscillations, but will fall off more quickly due to the complex exponential dependence that results in an inverse frequency contribution.  Those that are off-resonance are filtered out and do not appreciably change the probability.  Tuning the driving based on the RWA and CR terms suggests novel waveforms for sensing and the extraction of parameters in two-level systems.

\begin{figure*}[t]
    \centering
    \includegraphics[width=0.8\textwidth]{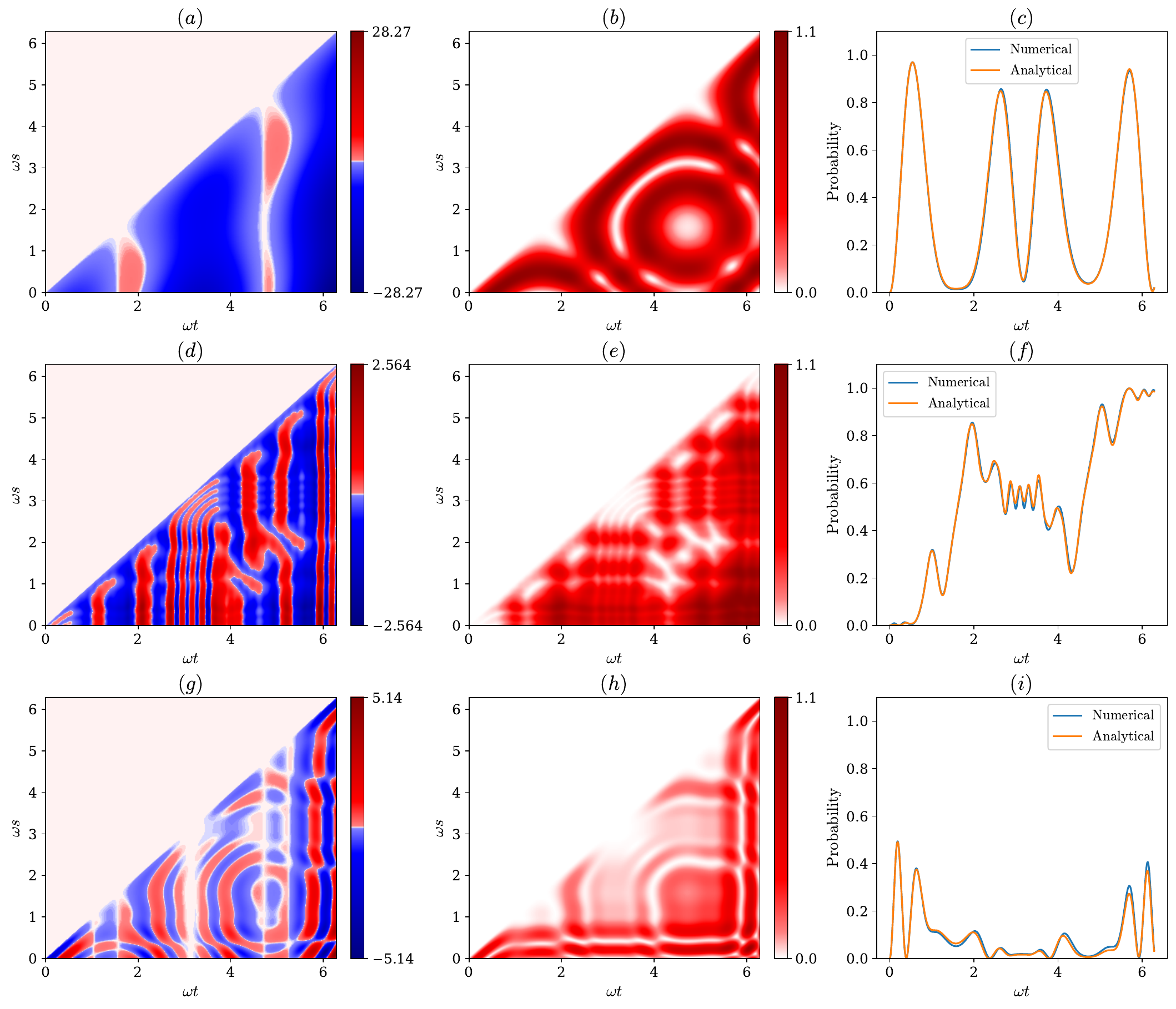}
    \caption{\textbf{Exact Transition Probability}:  Plots of the real part of the two-time kernel, two-time transition probability and the transition probability when $s=0$ for two differing waveforms in the interval $[0,T] \cup [0,T]$ for period $T=2\pi/\omega$.  In (a)-(c), the typical Bloch-Siegert Hamiltonian was used with energy bias $\epsilon_0 = \omega$ and amplitudes $\Delta_1 = 3\omega$.  In (d)-(f), the waveform used is a longitudinal modulation given by the first and third harmonics with amplitudes $A_1 = 13\omega$ and $A_2 = 18\omega$ with $\epsilon_0 = \omega$ and $\Delta = 1.5\omega$.  Lastly, (g)-(i) display the results for a longitudinal driving given by the first harmonic with amplitude $A_1 = 13\omega$ and transverse driving given by the first, second, and third harmonics with amplitudes $\Delta_i = 1.5\omega$ and constant energy bias and tunneling given by $\epsilon_0 = \omega$ and $\Delta = 1.5\omega$, respectively.  With more complex modulating waveforms, the two-time kernel and the two-time probability display rich interference patterns with specific features reflected in the single time variable probability (that is when $s=0$).}
    \label{fig:wide}
\end{figure*}

\subsection{Transition Probability}

Based on the unitary, we are able to determine the exact transition probability as $p(t,s)=|U_{12}(t,s)|^2$.  In Figure 2, we plot contours of the two-time kernel and the two-time transition probability alongside the probability with $s=0$ for the system driven beyond typical perturbative regimes.  Higher $\star$-powers of the Neumann expansion terms were efficiently computed using the relationship between the $\star$-product and matrix multiplication \cite{giscard2023exact}.  The analytical formulation matches well with numerical computations performed using \textit{QuTiP} \cite{johansson2012qutip}.  In particular, in Figure 2(a)-2(c), we plot the kernel and probability for the typical Bloch-Siegert Hamiltonian.  Figure 2(d)-2(f) displays the kernel and probability for a Hamiltonian given by a system driven by the first and third harmonic in the longitudinal direction.  Finally, Figure 2(g)-2(i) displays the kernel and probability for a system driven by the first and third harmonic in the longitudinal direction as well as the first, second, and third harmonics in the transverse direction.  The kernels display the defining features of a non-autonomous system, as mentioned previously in Figure 1.  Observations of the two-time probabilities reveal distinct features that appear in the two-time kernels.  In particular, general features of the Bloch-Siegert Hamiltonian in Figures 2(a)-2(c) and generalizations of this Hamiltonian with an increased number of harmonics include a number of contour bands that extend and terminate at the boundary, $t=s$.  The two-time probability appears to radiate out from a center that corresponds to a point in the two-time kernel that exhibits a negligible amplitude in the oscillations.  Similarly, general features of Hamiltonians with longitudinal modulation may be determined by observations of Figures 2(d)-2(f).  A defining feature of longitudinal modulated systems are the contour bands that repel away from the boundary, $t=s$.  The number of bands increases with the number of harmonics or the amplitude of the driving with the two-time probability maintaining these features.  When $s=0$, the transition probability maintains the contour bands as oscillations centered at some constant value. This value changes with the formation or extinction of the contour bands in the kernel, as observed for the region given between $\omega t = 4$ and $\omega t =6$ in Figure 2(d).  That is, in the region $\omega t =[2.5, 4]$ the contour bands correspond to five oscillations around the probability of approximately $p(t) = 0.58$ with the formation of a new band occurring at $t=4$, adding an additional oscillation close to $p(t) = 0.58$ and signaling for a transition to a new stop value for the probability.  Finally, when the modulation is included in both the longitudinal and transverse directions, the kernel reflects a complex interference pattern radiating out from a point in the two-time domain.  A general observation for the probability at $s=0$ with modulations of this form include the constructive interference at the boundaries leading to large peaks near $\omega t = 0$ and $ \omega t = T$ with a central region given by a negligible probability.

One may determine the average transition probability from the exact expression.  To perform the complex multiplication between divided differences one may use a divided difference Leibniz rule \cite{de2005divided}.  Or, appealing to the Hermite-Genocchi formula and Omega calculus, the products of individual divided differences are instead products of multivariate integrals over a polytope \cite{robins2021friendly}.    Instead, we note that the expansions are in a generalized exponential divided difference basis commonly found in the literature involving Riesz bases for controllability of string systems \cite{ullrich1980divided,avdonin2001exponential}.  Motivated by this, generalized divided difference bases of complex exponentials may be re-written in the basis $\left\lbrace e^{i\omega_l t},te^{i\omega_l t},\dots,t^K e^{i\omega_l t} \right\rbrace$ for some integer $K$ and frequency $\omega_l$.  Re-summing the series of divided differences results in a new series solution expanded in this basis.  For example, assuming the initial time variable is $s=0$ and denoting the product formula as 

\begin{equation}
\Pi_p = \prod_{j \neq p} (x_p - x_j) \textit{.}
\end{equation}

\noindent Then, the multivariate derivative of the divided differences involves summations of the following

\begin{multline}
\frac{1}{k_0!...k_n!} \frac{\partial^{k_0}}{\partial x_0^{k_0}}...\frac{\partial^{k_n}}{\partial x_n^{k_n}} \left(\frac{e^{i x_p t}}{\Pi_p}\right) = \\ 
\sum_{l=0}^p \binom{k_p}{l} (it)^{k_p} e^{i x_p t} \frac{\partial^{k_p-l}}{\partial x_p^{k_p-l}} \left( \prod_p^m \frac{k_i!}{(x_p-x_i)^{k_i}} \frac{1}{\Pi_p} \right) \textit{.}
\end{multline}

\noindent With re-indexing, all terms will result in terms of the form $t^K e^{ilt}$ or $t^K e^{i(\epsilon_0+l\omega)t}$ for some integer $K$ and some integer $l$.  The significant contributions for long term dynamics are those that correspond to the power of the time variable which arise from repeated differentiation of the complex exponential function.  The maximum repeated argument results in the highest power of time.  Once all coefficients have been determined, the series of the unitary may be written as

\begin{subequations}
\begin{equation}
U_{11} = \sum_{k=0}^\infty \sum_{l=-\infty}^\infty C_{kl} t^{k} e^{i l \omega t} + \sum_{k=0}^\infty \sum_{l=-\infty}^\infty D_{kl} t^{k} e^{i (\epsilon_0 + l \omega) t}\text{;}
\end{equation}
\begin{equation}
U_{21} = \sum_{k=0}^\infty \sum_{l=-\infty}^\infty Q_{kl} t^{k} e^{i l t} + \sum_{k=0}^\infty \sum_{l=-\infty}^\infty P_{kl} t^{k} e^{i (\epsilon_0 + l \omega) t} \textit{.}
\end{equation} 
\end{subequations}

To simplify the calculations here and in subsequent sections, we assume the energy bias is $\epsilon_0=-\alpha \omega$ for $\alpha \in \mathbb{Z}$, resulting in updated coefficients as $D_{kl}$($P_{kl}$) are re-indexed to $D_{k(l+\alpha)}$($P_{k(l+\alpha)}$).  After consolidation of the coefficients, the components of the unitary matrix are  

\begin{subequations}
\begin{equation}
U_{11} = \sum_{k=0}^\infty \sum_{l=-\infty}^\infty C_{kl} t^{k} e^{i l \omega t} \text{;}
\end{equation}
\begin{equation}
U_{21} = \sum_{k=0}^\infty \sum_{l=-\infty}^\infty Q_{kl} t^{k} e^{i l \omega t} \text{.}
\end{equation}
\end{subequations}

Once obtained, the transition probability may be found by multiplying the corresponding unitary components as a Cauchy product where one would require to properly re-index the coefficients subject to the repeated arguments of the divided difference.  The corresponding transition probability is

\begin{subequations}
\begin{equation}
p(t) = \sum_{k=1}^\infty \sum_{l=-\infty}^\infty O_{kl} t^{k} e^{i l \omega t}\text{;}
\end{equation}
\begin{equation}
O_{kl}=\sum_{q=1}^\infty \sum_{p=-\infty}^\infty \overline{Q}_{(p-k)(l-q)}Q_{kl} \text{.}
\end{equation}
\end{subequations}

\noindent The averaged transition probability is determined by integrating Eq. (30). 

As an example, the simplest coefficients are determined when all of the arguments of the exponential divided difference are equivalent.  This would correspond to all of the arguments being exactly zero, corresponding to resonance.  Assuming that the CR and OR terms may be ignored then the coefficients are

\begin{subequations}
\begin{equation}
C_{(2k)0} = \overbrace{\overline{\mathcal{J}_\alpha^\Delta} \dots \overline{\mathcal{J}_\alpha^\Delta}}^{\textit{k}} \overbrace{{\mathcal{J}_\alpha^\Delta} \dots \mathcal{J}_\alpha^\Delta}^{\textit{k}} \frac{i^{2k}}{2k!} \text{;}
\end{equation}
\begin{equation}
Q_{(2k+1)0} = \overbrace{\overline{\mathcal{J}_\alpha^\Delta} \dots \overline{\mathcal{J}_\alpha^\Delta}}^{\textit{k}} \overbrace{\mathcal{J}_\alpha^\Delta \dots \mathcal{J}_\alpha^\Delta}^{\textit{k+1}} \frac{i^{2k+1}}{(2k+1)!} \text{.}
\end{equation}
\end{subequations}

\noindent However, substituting the coefficients back into Eq. (29) results in nothing more than the Taylor series for the sine and cosine functions resulting in the solution of the Schr\"odinger equation and the transition probability in a perturbative regime.

\subsection{Floquet Eigenenergies and Effective Hamiltonians}

Lastly, as a consequence of the above procedure, we comment on the Floquet Hamiltonian and effective Hamiltonians generated by the periodic driving.  As investigated previously in the literature, Floquet theory will result in the dynamics of this system when performing numerical computations.  Floquet theory separates the unitary dynamics into a periodic contribution and a contribution from a non-periodic monodromy matrix such that the unitary can be factored into a periodic matrix and a monodromy matrix, $\mathcal{F}$, resulting in

\begin{equation}
U(t,0) = P(t)\mathcal{F} \textit{.}
\end{equation}

\noindent However, the monodromy matrix can be written as an exponentiated Floquet Hamiltonian $\mathcal{F}=e^{-i H_F T}$.  Therefore, within our construction, the Floquet Hamiltonian may be determined by $H_F = i  \text{ln}(U(T,0))/T$.  The quasienergies of this Hamiltonian may be determined from the properties of the eigenvalues of the unitary after a single period of evolution.  Due to the unitarity, the eigenvalues are guaranteed to lie on the unit circle, resulting in $\lambda_{\pm}=Tr(U)/2 \pm i\sqrt{1-(Tr(U)^2)/4}$ for eigenvalues, $\lambda_{\pm}$.  The eigenenergies of the Floquet Hamiltonian are $\lambda_{\pm} = e^{\pm i \epsilon T} = \cos(\epsilon T) \pm i \sin(\epsilon T)$ and obtained by the expression

\begin{multline}
\cos(\epsilon T) = 1 + \sum_{k=1} \sum_{\left\lbrace m,n \right\rbrace} \overline{\mathcal{J}_{\left\lbrace n \right\rbrace}^\Delta} \mathcal{J}_{\left\lbrace m \right\rbrace}^\Delta \\ \times \cos([\epsilon_0 + M_k - N_k, M_{k-1} - N_{k-1},\dots,\epsilon_0 + M_1, 0]T)\text{,}
\end{multline}

\noindent where the eigenenergies may be multivalued, as expected.  Determining the exact Floquet Hamiltonian would be challenging due to the logarithm.  Instead, an effective Hamiltonian may be determined directly from the Schr\"odinger equation over one period of evolution defined as

\begin{equation}
H_{eff} = \frac{i}{T} \int_0^T U^\dag(t) \frac{dU}{dt}  \, dt \textit{.}
\end{equation}

\noindent Differentiating and re-indexing the unitary components in Eq. (29),  

\begin{subequations}
\begin{equation}
\frac{dU_{11}}{dt} = \sum_{k=0}^\infty \sum_{l=-\infty}^\infty \tilde{C}_{kl} t^{k} e^{i l \omega t} \text{;}
\end{equation}
\begin{equation}
\frac{dU_{21}}{dt} = \sum_{k=0}^\infty \sum_{l=-\infty}^\infty \tilde{Q}_{kl} t^{k} e^{i l \omega t} \text{.}
\end{equation}
\end{subequations}

\noindent Similar in the procedure to determine the average transition probability, we perform a Cauchy product, then integrate.  The following identity may be used when $c \ne 0$

\begin{equation}
\int_0^T t^n e^{ict} \, dt = e^{i c T} \sum_{j=0}^{n} (-1)^{n-j} \frac{n!}{j!(ic)^{n-j+1}}T^j \text{.}
\end{equation}

\noindent The effective Hamiltonian is 

\begin{subequations}
\begin{equation}
H(t)= \frac{1}{2} \begin{pmatrix}
H_{11} & H_{12} \\ 
H_{21} & H_{22}
\end{pmatrix} \text{;}
\end{equation}
\begin{multline}
H_{11} = \sum_{k=1}^\infty  \frac{H_{kl}^{11}}{k} T^{k} \\ + \sum_{k=1}^\infty \sum_{l=-\infty}^\infty \sum_{j=0}^{k-1} (-1)^{k-j} \frac{H_{kl}^{11} (k)!}{j!(il\omega)^{k-j+1}}T^j \text{;}
\end{multline}
\begin{multline}
H_{12} = \sum_{k=1}^\infty  \frac{H_{kl}^{12}}{k} T^{k} \\ + \sum_{k=1}^\infty \sum_{l=-\infty}^\infty \sum_{j=0}^{k} (-1)^{k-j} \frac{H_{kl}^{12} (k)!}{j!(il\omega)^{k-j+1}}T^j \text{;}
\end{multline}
\begin{equation}
H_{kl}^{11} = \sum_{k=0}^\infty \sum_{l=-\infty}^\infty Q_{(k-m)(n-l)}\tilde{C}_{mn} + \tilde{Q}_{(k-m)(n-l)}C_{mn} \text{;}
\end{equation}
\begin{equation}
H_{kl}^{12} = \sum_{k=0}^\infty \sum_{l=-\infty}^\infty \tilde{C}_{(k-m)(n-l)}C_{mn} + \tilde{Q}_{(k-m)(n-l)}Q_{mn} \text{'}
\end{equation}
\end{subequations}

\noindent where $H_{21}=\overline{H_{12}}$ and $H_{22}=-H_{11}$.  The resulting expressions are expansions in terms of the period, or inverse frequency, as expected.  

\section{Perturbative Regime}

Any truncation of the kernel will result in perturbative solutions to the transition probability, however dominant terms of the kernel may be retained for a transition probability that closely matches that of the exact probability.  For example, assuming weak transverse terms, we ignore the CR and OR terms in Eq (23).  Then, the remaining term is the RWA term.  Carrying out the $\star$-resolvent calculation and the integration for the unitary components, the resulting series will be the Taylor series of the sine and cosine functions. Of course, the effective Hamiltonian matches what would be determined from the RWA when observing the Hamiltonian in Eq. (13).  For example, the resulting effective Hamiltonian from our exact expression in Eq. (37), or performing the RWA as is commonly performed results in 

\begin{equation}
H_{RWA}=  \begin{pmatrix}
0 & \mathcal{J}_l^\Delta \\ 
\overline{\mathcal{J}_l^\Delta} & 0
\end{pmatrix}\textit{.}
\end{equation}

\noindent From the Hamiltonian the generalized Rabi frequency is $|\mathcal{J}_l^\Delta|$, corresponding to the GBFs, and would be useful in amplitude spectroscopy applications \cite{satanin2014amplitude,satanin2012amplitude,berns2008amplitude}.

We plot the Rabi frequencies for various waveforms in Figure 3.  Due to the difficulty of visualizing interference patterns beyond two dimensions, we only plot interference patterns associated with the longitudinal modulation having two harmonics with the axes given by the amplitudes of those harmonics.  For Figures 3(a)-3(c), we plot the Rabi frequency with the first and second harmonic for longitudinal modulation.  Figure 3(a) includes a constant energy bias given by $\epsilon_0 \approx \omega$.  Figure 3(b) includes a constant energy bias given by $\epsilon_0 \approx 10\omega$.  Finally, Figure 3(c) includes a constant energy bias given by $\epsilon_0 \approx 10\omega$ and includes transverse modulation given by the first and twentieth harmonics.  For Figures 3(d) - 3(f) we plot the Rabi frequency for transverse modulations given by the third and eleventh harmonic.  One may observe in plots 3(a) and 3(b) that there are differing regions of behavior separated by bifurcation surfaces.  For example, in Figure 3(a), the regions on the upper and lower ends of the plots, intersecting the y-axis, consist of discrete, closed loops, while the contours intersecting the x-axis are extended, curved loops with symmetry about the y-axis.  Similar patterns with biharmonic driving were observed in \cite{satanin2014amplitude} and reproduced here with our method.  The looping structures are maintained when considering Rabi frequencies dependent on higher order GBFs corresponding to larger energy biases, as seen in Figure 3(b), where an opening at the origin begins to develop.  This region exhibits exponential decay where the magnitude of the GBF and the corresponding Rabi frequency is negligible.  Though it is difficult to discern a general structure consisting of discrete loops for figures 3(d) and 3(e), similar observations can be made at larger amplitudes than those plotted.  However, the general structure consists of a large region of exponential decay that extends along the y-axis and is reminiscent of the modal structure that would be obtained if one were to consider longitudinal modulations with incommensurate frequencies.  This similarity is due to the rationality between the third and eleventh harmonic in the driving waveform.  Lastly, Figures 3(c) and 3(f) result in complex structure of overlapping interference patterns.  The analytical description of the topology of the interference patterns may be decomposed into bifurcation surfaces that clearly separate the regions of the GBFs \cite{kuklinski2019identities}.  For those encountered in Figures 3(a), 3(b), 3(c), and 3(d), the algebraic surfaces involve bifurcation lines and ellipses to denote the differing dynamical regions of the GBFs.  However, with the addition of the transverse modulations, the number of bifurcation surfaces increase with the surfaces overlapping resulting in complex separations in the interference regions associated with the Rabi frequency.  In terms of Floquet theory, the Rabi frequency is proportional to the Floquet quasienergy gaps, therefore, the topology given by the zeros of the GBFs dictate the closing of these gaps and the coherent destruction of tunneling.

\begin{figure*}[t]
    \centering
    \includegraphics[width=0.8\textwidth]{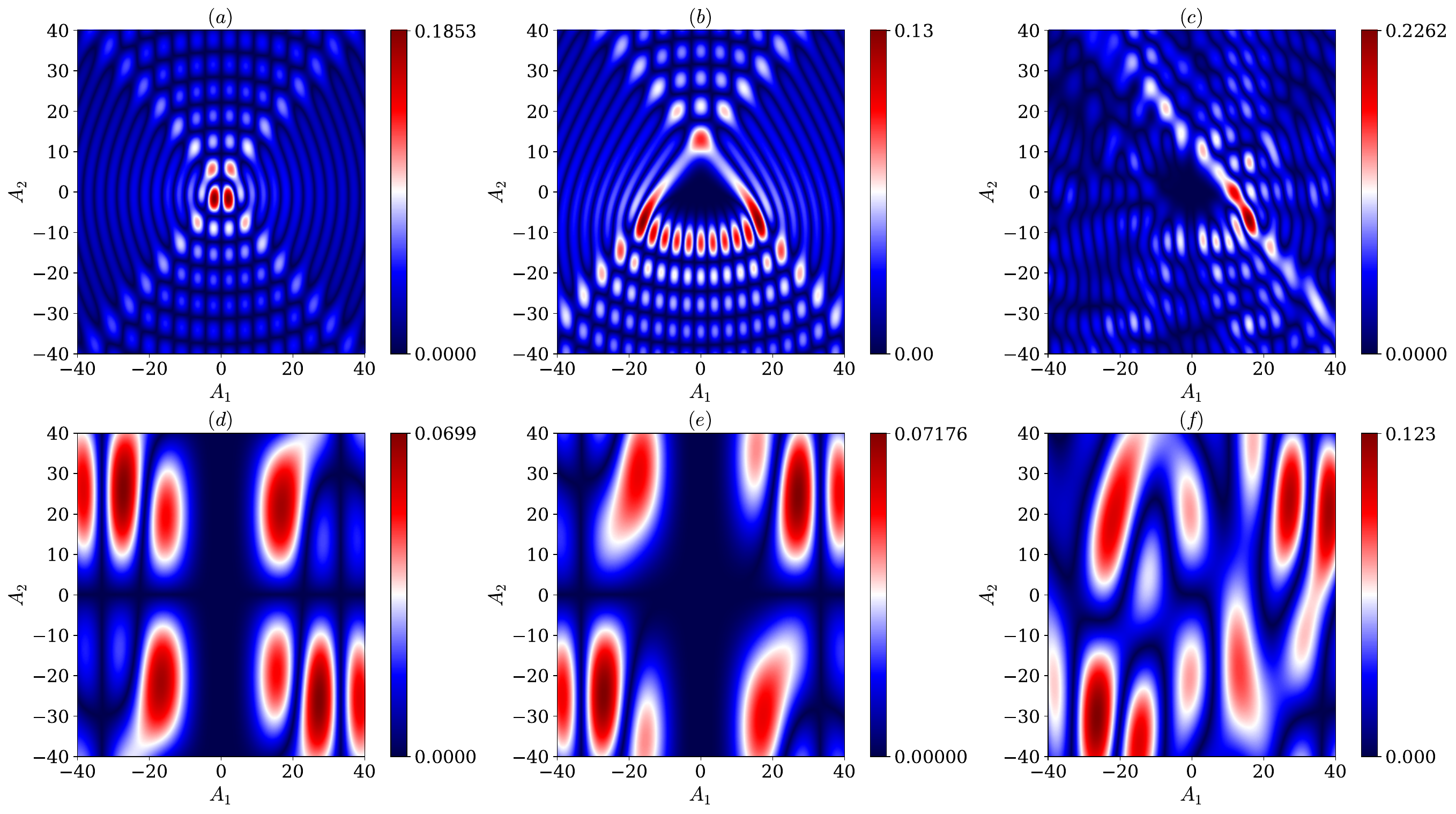}
    \caption{\textbf{Generalized Rabi Frequencies}:  Plots of the generalized Rabi frequencies $|\mathcal{J}_l^\Delta (A_1, A_2)|$ for $-40\omega<A_1,A_2<40\omega$ for various waveforms.  The waveforms used are (a) first and second harmonics in the diagonal with a DC bias strength of $\epsilon_0 \approx \omega$ , (b) first and second harmonics in the diagonal with a DC bias strength of $\epsilon_0 \approx 10\omega$, (c) first and second harmonics in the diagonal with a DC bias strength of $\epsilon_0 \approx 10 \omega$ with transverse modulation given by the first and twentieth harmonics, (d) the third and eleventh harmonic in the diagonal with a DC bias strength of $\epsilon_0 \approx \omega$, (e) the third and eleventh harmonic in the diagonal with a DC bias strength of $\epsilon_0 \approx 10\omega$, and (f) the third and eleventh harmonic in the diagonal with a DC bias strength of $\epsilon_0 \approx 10 \omega$ with transverse modulation given by the first and twentieth harmonics.  Note here for (a) the Rabi frequencies pick up the GBF structure in the plots with a radially divergent structure as seen in \cite{satanin2014amplitude,kuklinski2019identities} reproduced using our method here.  Plot (b) maintains this divergent structure with an ever increasing region of exponential decay associated with higher order GBFs.  Plots (d) and (e) seem to remove this divergent structure and appear to be similar in structure, however, if one were to expand the amplitudes to larger magnitudes, the similarity in the structure would vanish and would result in dynamics clearly separated similar to (a) and (b).  Lastly, when the longitudinal modulation is turned on, the Rabi frequencies change drastically with overlapping constructive/destructive interference patterns.}
    \label{fig:wide}
\end{figure*}

The resulting transition probability in time and the time averaged probability are

\begin{subequations}
\begin{equation}
p(t) = \frac{1}{2}\sum_{l=-\infty}^{\infty} \frac{(\mathcal{J}_l^\Delta)^2}{\Omega_l^2} \left(1-\cos\left(\Omega_l t\right) \right) \text{;}
\end{equation}
\begin{equation}
p_{ave} = \frac{1}{2}\sum_{l=-\infty}^{\infty} \frac{(\mathcal{J}_l^\Delta)^2}{\Omega_l^2} \textit{;}
\end{equation}
\begin{equation}
\Omega_l = \sqrt{(\mathcal{J}_l^\Delta)^2+(l\omega - \epsilon_0)^2}
\end{equation} \text{.}
\end{subequations}

\noindent In Figure 4, the time-dependent and the averaged transition probability is plotted and compared to numerical computations using \textit{QuTiP}.  In the regime where the coupling strength remains sufficiently low, the analytical results agree well with the numerical results.  In Figures 4(a) and 4(b), we plot the probability of a transverse modulation given by the first and second harmonic with a constant energy bias given by $\epsilon_0 \approx \omega$.  In Figures 4(d) and 4(e), we plot the average probability of a transverse modulation given by the third and eleventh harmonic with a constant energy bias of $\epsilon_0 \approx \omega$ and transverse modulation given by the first and twentieth harmonic.  Similar to 4(c), we plot the time dependent probability for the third and eleventh harmonic in 4(f).  Finally, for Figures 4(g)-4(i) we plot the average probability and time dependent probability for a longitudinal modulation given by the first and second harmonics with transverse modulation given by the first, third, fourth, seventh, ninth, tenth, twelfth, thirteenth, fifteenth, seventeenth, eighteenth, and twentieth harmonic.  Observing the amplitude dependence of the time-averaged probability, we note that, while the widths of the resonances differ between the analytical and numerical probability, the interference patterns and topology of the GBFs noted in the above discussion for the Rabi frequency persist for the time-averaged probability.  It is clear that the perturbative regime does well for the time dependent probability as can be seen in Figures 4(c) and 4(f).  In Figures 4(g)-4(i), the overall interference pattern remains however, the deviation from the perturbative regime begins to become more pronounced.  This is also encountered for long times for the time-dependent probability, requiring our expression for the exact transition probability.

\begin{figure*}[t]
    \centering
    \includegraphics[width=0.8\textwidth]{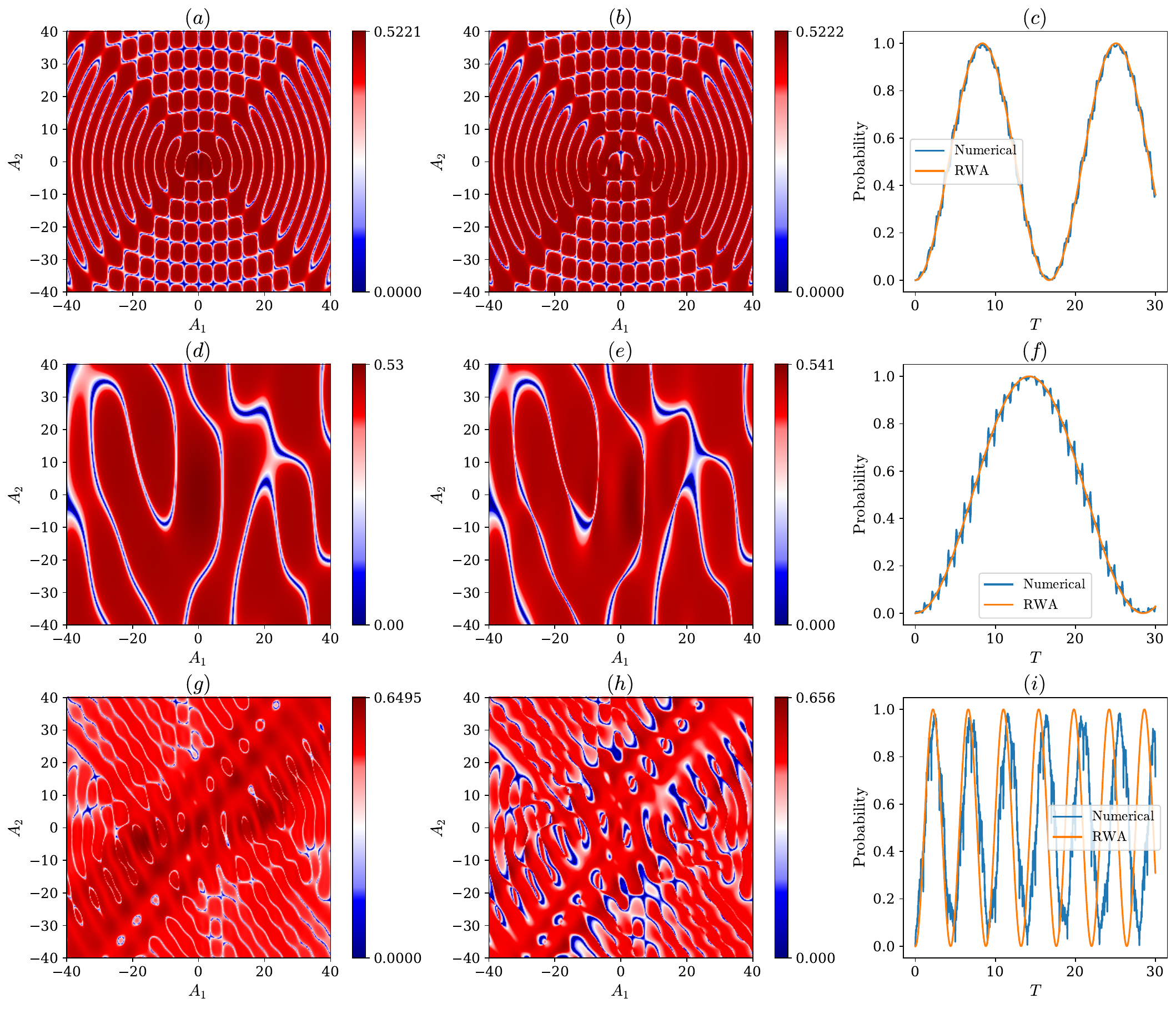}
    \caption{\textbf{Perturbative Transition Probability from Truncated Kernel}: Analytical (left column) and numerical (middle column) plots of the time-averaged upper level occupation probabilities $p(A_1, A_2)$ for $-40\omega<A_1,A_2<40\omega$ with the corresponding time-dependent probabilities displayed in the right column.  The waveforms used for (a)-(c) are the first and second harmonics in the diagonal with a DC bias strength of $\epsilon_0 \approx \omega$ and a tunneling strength of $\Delta = 0.25 \omega$, (d)-(f) are the third and eleventh harmonics in the diagonal with a DC bias strength of $\epsilon_0 \approx \omega$ and a tunneling strength of $\Delta = 0.25 \omega$ with transverse modulation given by the first and twentieth harmonics with amplitudes $\Delta_i = 0.25 \omega$ and (g)-(i) are the first and second harmonics in the diagonal with a DC bias strength of $\epsilon_0 \approx \omega$ and a tunneling strength of $\Delta = 0.25 \omega$ with transverse modulation given by the first, third, fourth, seventh, ninth, tenth, twelfth, thirteenth, fifteenth, seventeenth, eighteenth, and twentieth harmonics with amplitudes $\Delta_i = 0.5 \omega$.  We call attention to the accuracy of resonance positions with our analytical treatment using GBFs compared to numerical computations.  Furthermore, the magnitudes of the occupation probabilities match well.  However, with increasing transverse modulation strength and increasing the number of harmonics, the resonance widths vary greatly from the analytical description.  Furthermore, the long time dynamics deviates from the approximation regime, as expected.}
    \label{fig:wide}
\end{figure*}

\section{Discussion and Applications}

Our results provide an analytical framework for investigating the transition probabilities of a periodically driven two-level system under transverse or longitudinal modulations with any number of harmonics.  Specifically, it provides an alternative avenue for the investigation of modulated two-level systems beyond the Floquet formalism, other numerical methods, and perturbative regimes.  While the analysis here is restricted to modulations given by finite sums of harmonically related waveforms, the method could, in principle, be extended to non-harmonically related and/or infinite waveforms \cite{lorenzutta1997infinite,bekers2024extending}.   A more thorough investigation of the coefficients of the transition probability and the effective Hamiltonian may be of interest to determine analytical corrections to the Rabi frequency and the dynamics of the two-level system without using high frequency methods, such as the Magnus expansion.  For example, an additional investigation may result in novel analytical expressions for generalized Bloch-Siegert shifts that arise with additional harmonics included in the transverse modulation.  Furthermore, in this investigation, the kernel was separated based upon the resonant dynamics of the system, however, kernel accelerations to remove or promote components of the kernel will result in novel analytical expressions that may reveal features of the dynamics hidden in the general expression given in Eqs. (16) or (17) \cite{giscard2020solutions}. Similarly, combining the analytical method presented here with novel frame changes \cite{giscard2025novel} offers a non-perturbative avenue for investigating dynamical resonances that appear in quantum materials, NMR, or coupled qubit systems \cite{gomez2023anomalous,jurkutat2025nuclear,jeschke2025unexpected,kropf2012nonsecular,sauer2012entanglement} 

Beyond the theoretical investigations that may arise using this method, these findings suggest immediate applications to quantum technologies.  A natural consequence of the discrete parameters as the arguments of the GBFs in sonar/radar applications is the ability to determine optimization algorithms.  For example, Felton and Hague, developed optimization algorithms synthesizing waveforms with low Auto-Correlation Function sidelobes \cite{felton2023gradient}.  A natural extension of this work would be to determine optimization algorithms for the dynamics of a two-level system.  For example, in a quantum sensor, an optimal control drive may be tailored around the signal of interest by separating the kernel into a control and a signal kernel and investigating the resulting dynamics, seeking the waveform that maximizes the signal-to-noise ratio.  Or, using the kernel and Volterra series methods offer opportunities for novel signal processing algorithms for the extraction of parameters contained in the signal \cite{mirri2002modified}.  Alternatively, for quantum gates, novel control algorithms may be implemented based on the kernel expression provided and the dynamics in the two-time domain.  For example, stitching together unitaries derived from differing kernels may be advantageously used to realize higher fidelity gates.

\section{Conclusion}

In conclusion, we introduced the multi-tone sinusoidal frequency modulation (MTSFM) to a two-level quantum system through the modulation of the energy bias and the tunneling strength.  We introduce the generalized Bessel functions (GBFs) to consolidate the modulations to a single term.  To find the evolution and transition probability we derive exact analytical expressions for the unitary generated from a two-time kernel that contains the GBFs and all of the information associated with the driving.  The advantage of this exact approach in comparison to perturbative approaches lies in the kernel expression.  In an application, such as quantum sensing, the kernel expression will contain the signal and control kernel for waveform optimization of the signal-to-noise.  Similarly, novel maximum likelihood estimation algorithms using the kernel would allow for the extraction of relevant signal parameters from a quantum sensor.

\section{Acknowledgements}

This research was funded by the Naval Undersea Warfare Center  In-House Laboratory Independent Research (ILIR) program from the Office of Naval Research (ONR) under N0001424WX00177.  
V.F.M. acknowledges funds from Brown University. 
The author would   like to thank P.L. Giscard for helpful discussions on the solutions of non-autonomous differential equations and the $\star$-resolvent formalism. 
We also thank J. B. Marston, D. Feldman, A. Rosuel, and A Hui    for valuable  reviews of the manuscript. 

\appendix

\section{Calculation of Eq. (21)}

We provide a non-rigorous example calculation of Eq. (21) from the main text.  We require several theorems and identities from prior works \cite{kalev2021integral,baxterexponential}.  For convenience we state the theorems and identities here, without proof.  

\begin{theorem}[Hermite-Genocchi, \cite{baxterexponential}]

Let $f \in C^{(n)}(\mathbb{R})$ and let $a_0, a_1, \dots, a_n$ be (not necessarily distinct) real numbers.  Then, for $n \geq 1$,

\begin{multline}
f[a_0,a_1,\dots,a_n] = \\ \int_{S_n} f^{(n)}(t_0 a_0 + t_1 a_1 + \dots + t_n a_n) \, dt_1 \dots dt_n \\
= \int_0^1 dt_1 \int_0^{1-t_1} dt_2 \dots \int_0^{1-\sum_{k=1}^{n-1} t_k} \\ \times f^{(n)}(t_0 a_0 + t_1 a_1 + \dots + t_n a_n) \text{,}
\end{multline}

where the domain of integration is the simplex

\begin{equation}
S_n = \left\lbrace t = (t_1, t_2, \dots, t_n) \in R_+^n: \sum_{k=1}^n t_k \leq 1 \right\rbrace
\end{equation}

and

\begin{equation}
t_0 = 1 - \sum_{k=1}^n t_k \text{.}
\end{equation}

\end{theorem}

\noindent Next, we require a generalization of the Hermite-Genocchi theorem.

\begin{theorem}[\cite{baxterexponential}]

Let $V \in \mathbb{R}^{(n \times n)}$ be any nonsingular matrix.  Then

\begin{equation}
\frac{1}{|\text{det}V|} \int_{K(V)} f^{(n)}(a^T y)dy = f[0,(V^T a)_1, \dots, (V^T a)_n]\text{,}
\end{equation}

\noindent where $(V^T a)_k$ denotes the $k$-th component of the vector $V^T a$.

\end{theorem}

\noindent Finally, an identity that has proven useful in \cite{kalev2021integral,kalev2021quantum,chen2021quantum} is

\begin{identity}[\cite{kalev2021integral}]

\begin{equation}
(i)^q \int_0^t dt_q \dots \int_0^{t_2} dt_1 e^{i\gamma_q t_q + \dots + \gamma_1 t_1} = e^{it[x_0,x_1,\dots,x_{q-1},0]}
\end{equation}

\noindent where $x_j = \sum_{k=j+1}^q \gamma_k$. 
\end{identity}

\noindent Note that Theorem 1 and Identity 1 are related through Theorem 2.  That is, Theorem 2 encapsulates Identity 1 when $t=1$ with a vertex matrix given by an upper triangular matrix of ones.  The statement of Identity 1 is significant for this work as it relates the integral bounds for the integration region from the unit simplex to a simplex contained within a region bounded by the time variable, $t$.  Within the subsequent calculation we find that each additional $\star$-power expands the region of integration to a higher dimensional simplex, allowing for the use of Theorem 2 and Identity 1 together to define the vertex matrix corresponding to this new simplex, immediately providing a representation within the divided difference calculus.  

Proceeding with this calculation, we re-state the kernel here, 

\begin{equation}
K(t,s) = (-i) \sum_{mn} \overline{\mathcal{J}_n^\Delta} \mathcal{J}_m^\Delta e^{i (m\omega - n \omega) t} e^{i[f_n,0](t-s)} \text{.}
\end{equation}

\noindent We first perform the computation for the second $\star$-power. Then,

\begin{multline}
(K \star K)(t,s) = (-i)^2 \sum_{mnlk} \overline{\mathcal{J}_n^\Delta} \mathcal{J}_m^\Delta \overline{\mathcal{J}_k^\Delta} \mathcal{J}_l^\Delta \\ 
\times \int_s^t e^{i (m\omega - n \omega) t} e^{i[f_n,0](t-\tau)} e^{i (l\omega - k \omega) \tau} e^{i[f_k,0](\tau-s)} \, d\tau \text{.}
\end{multline}

\noindent For convenience, we suppress the summation and the GBFs in subsequent calculations.  After performing a change of variables and expanding the divided difference terms using Identity 1, our integral expression may be written as

\begin{multline}
e^{i (m\omega - n \omega + l\omega - k\omega) t} \\
\times \int_0^{t-s} e^{i[f_n,0](u)} e^{-i (l\omega - k \omega)u} e^{i[f_k,0](t-s-u)} \, du \\
= e^{i (m\omega - n \omega + l\omega - k\omega) t} \\ \times  \int_0^{t-s} \int_0^{u} \int_0^{t-s-u} e^{i f_n v -i (l\omega - k \omega)u +i f_k y} \, du \, dv \, dy \text{.}
\end{multline}

\noindent Recognizing that the integral bounds are linearly related and appealing to Theorem 2, we are able to define a region bounded by a tetrahedron with vertices, in matrix form, given by

\begin{equation}
V=  \begin{pmatrix}
1 & 1 & 0 \\ 
0 & 1 & 0 \\
0 & 0 & 1 \\
\end{pmatrix}
\end{equation}

\noindent and coefficients given by

\begin{equation}
a =  \begin{pmatrix}
-l\omega + k\omega \\ 
f_n \\
f_k \\
\end{pmatrix} \text{.}
\end{equation}

\noindent By Theorem 2, the resulting divided difference expression is 

\begin{multline}
(K \star K)(t,s) = (-i)^3 \sum_{mnlk} \overline{\mathcal{J}_n^\Delta} \mathcal{J}_m^\Delta \overline{\mathcal{J}_k^\Delta} \mathcal{J}_l^\Delta e^{i (m\omega - n \omega + l\omega - k\omega) t} \\ \times e^{i[(V^T a)_1, (V^T a)_2, (V^T a)_3, 0](t-s)} \\ = (-i)^3 \sum_{mnlk} \overline{\mathcal{J}_n^\Delta} \mathcal{J}_m^\Delta \overline{\mathcal{J}_k^\Delta} \mathcal{J}_l^\Delta e^{i (m\omega - n \omega + l\omega - k\omega) t} \\ \times e^{i[ \epsilon_0 + n\omega -k\omega +l\omega,-k\omega +l\omega, \epsilon_0 + k\omega, 0](t-s)} \text{.}
\end{multline}

\noindent Where in the last step we permute the arguments of the divided difference because the divided difference is symmetric under any permutation of its arguments. 

Continuing this procedure, the third $\star$-power, after a change of variables, will result in an integral expression 

\begin{multline}
e^{i (m\omega - n \omega + l\omega - k\omega + p \omega - q\omega) t} \\ \times \int_0^{t-s} e^{i[ \epsilon_0 + n\omega -k\omega +l\omega,-k\omega +l\omega, \epsilon_0 + k\omega, 0](u)} \\ \times e^{-i (p\omega - q \omega)u} e^{i[f_q,0](t-s-u)} \, du
\end{multline}
\begin{multline}
=e^{i (m\omega - n \omega + l\omega - k\omega + p \omega - q\omega) t} \\ \times \int_0^{t-s} du \int_0^{u} du_1 \int_0^{u_1} du_2 \int_0^{u_2} du_3 \int_0^{t-s-u} dv \\ \times e^{i (l\omega - k\omega) u + i f_k u_1 - i f_l u_2 + f_n u_3 + i f_q v} \text{.}
\end{multline}

\noindent With the vertex matrix

\begin{equation}
V=  \begin{pmatrix}
1 & 1 & 1 & 1 & 0 \\ 
0 & 1 & 1 & 1 & 0 \\
0 & 0 & 1 & 1 & 0 \\
0 & 0 & 0 & 1 & 0\\ 
0 & 0 & 0 & 0 & 1 \\
\end{pmatrix}
\end{equation}

\noindent and coefficients given by

\begin{equation}
a =  \begin{pmatrix}
-l\omega + k\omega \\ 
f_k \\
-f_l \\
f_n \\
f_q \\
\end{pmatrix} \text{,}
\end{equation}

\noindent we immediately obtain the expression

\begin{widetext}
\begin{multline}
(-i)^5 e^{i (m\omega - n \omega + l\omega - k\omega + p \omega - q\omega) t} e^{i[ \epsilon_0 + n\omega -k\omega +l\omega- q\omega +  p \omega,-k\omega +l\omega - q\omega +  p \omega, \epsilon_0 + k\omega - q\omega +  p \omega, - q\omega +  p \omega, \epsilon_0 - q\omega, 0](t-s)}
\end{multline}
\end{widetext}

Finally, the above procedure may be repeated, then the $k$-th $\star$-power is obtained as 

\begin{widetext}
\begin{multline}
e^{i(M_k-N_k)t} \int_0^{t-s} e^{i[\epsilon_0 - M_{k-2} + N_{k-1}, -M_{k-3} + N_{k-2}, \dots, \epsilon_0 + M_1, 0](u)} e^{-i (m_k \omega - n_k \omega)u} e^{i[f_{n_k},0](t-s-u)} \, du \\ =  e^{i(M_k-N_k)t} \int_0^{t-s} du \int_0^{u} du_{k} \dots \int_0^{u_{k-2}} du_{k-1} \int_0^{t-s-u} dv  e^{-i (m_{k-1} \omega - n_{k-1} \omega) u + i f_{m_{k-2}} u_1 - i f_{n_{k-2}} u_2 + \dots + f_{n_k} u_1 + i f_{m_k} v} \text{.}
\end{multline}
\end{widetext}

\noindent With vertex matrix

\begin{equation}
V=  \begin{pmatrix}
1 & 1 & \dots & 1 & 0 \\ 
0 & 1 & \dots & 1 & 0 \\
\vdots & \vdots & \ddots & \vdots & \vdots \\
0 & 0 & 0 & 1 & 0\\ 
0 & 0 & 0 & 0 & 1 \\
\end{pmatrix}
\end{equation}

\noindent and coefficients given by

\begin{equation}
a =  \begin{pmatrix}
-n_{k-1} \omega + m_{k-1} \omega \\ 
f_{n_{k-2}} \\
f_{m_{k-2}} \\
\vdots \\
f_{n_1} \\
f_{m_1} \\
\end{pmatrix} \text{,}
\end{equation}

\noindent we are able to obtain the $k$-th $\star$-power of the kernel expression in the main text using Theorem 2.

\bibliographystyle{unsrt} 
\bibliography{mybib} 

@article{eckardt2017colloquium,
	author = {Eckardt, Andr{\'e}},
	journal = {Reviews of Modern Physics},
	number = {1},
	pages = {011004},
	publisher = {APS},
	title = {{Colloquium: Atomic quantum gases in periodically driven optical lattices}},
	volume = {89},
	year = {2017}}

@article{rodriguez2021low,
	author = {Rodriguez-Vega, Martin and Vogl, Michael and Fiete, Gregory A},
	journal = {Annals of Physics},
	pages = {168434},
	publisher = {Elsevier},
	title = {{Low-frequency and Moir{\'e}--Floquet engineering: A review}},
	volume = {435},
	year = {2021}}

@article{goldman2015periodically,
	author = {Goldman, Nathan and Dalibard, Jean and Aidelsburger, Monika and Cooper, Nigel R},
	journal = {Physical Review A},
	number = {3},
	pages = {033632},
	publisher = {APS},
	title = {Periodically driven quantum matter: The case of resonant modulations},
	volume = {91},
	year = {2015}}

@article{sen2021analytic,
	author = {Sen, Arnab and Sen, Diptiman and Sengupta, K},
	journal = {Journal of Physics: Condensed Matter},
	number = {44},
	pages = {443003},
	publisher = {IOP Publishing},
	title = {Analytic approaches to periodically driven closed quantum systems: methods and applications},
	volume = {33},
	year = {2021}}

@article{weitenberg2021tailoring,
	author = {Weitenberg, Christof and Simonet, Juliette},
	journal = {Nature Physics},
	number = {12},
	pages = {1342--1348},
	publisher = {Nature Publishing Group UK London},
	title = {{Tailoring quantum gases by Floquet engineering}},
	volume = {17},
	year = {2021}}

@article{oka2019floquet,
	author = {Oka, Takashi and Kitamura, Sota},
	journal = {Annual Review of Condensed Matter Physics},
	number = {1},
	pages = {387--408},
	publisher = {Annual Reviews},
	title = {Floquet engineering of quantum materials},
	volume = {10},
	year = {2019}}

@article{grifoni1998driven,
	author = {Grifoni, Milena and H{\"a}nggi, Peter},
	journal = {Physics Reports},
	number = {5-6},
	pages = {229--354},
	publisher = {Elsevier},
	title = {Driven quantum tunneling},
	volume = {304},
	year = {1998}}

@article{silveri2017quantum,
	author = {Silveri, MP and Tuorila, JA and Thuneberg, EV and Paraoanu, GS},
	journal = {Reports on Progress in Physics},
	number = {5},
	pages = {056002},
	publisher = {IOP Publishing},
	title = {Quantum systems under frequency modulation},
	volume = {80},
	year = {2017}}

@article{ivakhnenko2023nonadiabatic,
	author = {Ivakhnenko, V and Shevchenko, Sergey N and Nori, Franco},
	journal = {Physics Reports},
	pages = {1--89},
	publisher = {Elsevier},
	title = {{Nonadiabatic Landau--Zener--St{\"u}ckelberg--Majorana transitions, dynamics, and interference}},
	volume = {995},
	year = {2023}}

@article{lindner2011floquet,
	author = {Lindner, Netanel H and Refael, Gil and Galitski, Victor},
	journal = {Nature Physics},
	number = {6},
	pages = {490--495},
	publisher = {Nature Publishing Group UK London},
	title = {Floquet topological insulator in semiconductor quantum wells},
	volume = {7},
	year = {2011}}

@article{wang2013observation,
	author = {Wang, YH and Steinberg, Hadar and Jarillo-Herrero, Pablo and Gedik, Nuh},
	journal = {Science},
	number = {6157},
	pages = {453--457},
	publisher = {American Association for the Advancement of Science},
	title = {{Observation of Floquet-Bloch states on the surface of a topological insulator}},
	volume = {342},
	year = {2013}}

@article{rudner2013anomalous,
	author = {Rudner, Mark S and Lindner, Netanel H and Berg, Erez and Levin, Michael},
	journal = {Physical Review X},
	number = {3},
	pages = {031005},
	publisher = {APS},
	title = {{Anomalous Edge States and the Bulk-Edge Correspondence for Periodically<? format?> Driven Two-Dimensional Systems}},
	volume = {3},
	year = {2013}}

@article{tsuji2011dynamical,
	author = {Tsuji, Naoto and Oka, Takashi and Werner, Philipp and Aoki, Hideo},
	journal = {Physical Review Letters},
	number = {23},
	pages = {236401},
	publisher = {APS},
	title = {{Dynamical Band Flipping in Fermionic Lattice Systems: An ac-Field-Driven Change<? format?> of the Interaction from Repulsive to Attractive}},
	volume = {106},
	year = {2011}}

@article{gorg2018enhancement,
	author = {G{\"o}rg, Frederik and Messer, Michael and Sandholzer, Kilian and Jotzu,Gregor and Desbuquois, R{\'e}mi and Esslinger, Tilman},
	journal = {Nature},
	number = {7689},
	pages = {481--485},
	publisher = {Nature Publishing Group UK London},
	title = {Enhancement and sign change of magnetic correlations in a driven quantum many-body system},
	volume = {553},
	year = {2018}}

@article{heyl2018dynamical,
	author = {Heyl, Markus},
	journal = {Reports on Progress in Physics},
	number = {5},
	pages = {054001},
	publisher = {IOP Publishing},
	title = {Dynamical quantum phase transitions: a review},
	volume = {81},
	year = {2018}}

@article{zvyagin2016dynamical,
	author = {Zvyagin, AA},
	journal = {Low Temperature Physics},
	number = {11},
	pages = {971--994},
	publisher = {AIP Publishing},
	title = {Dynamical quantum phase transitions},
	volume = {42},
	year = {2016}}

@article{choi2017observation,
	author = {Choi, Soonwon and Choi, Joonhee and Landig, Renate and Kucsko, Georg and Zhou, Hengyun and Isoya, Junichi and Jelezko, Fedor and Onoda, Shinobu and Sumiya, Hitoshi and Khemani, Vedika and others},
	journal = {Nature},
	number = {7644},
	pages = {221--225},
	publisher = {Nature Publishing Group UK London},
	title = {Observation of discrete time-crystalline order in a disordered dipolar many-body system},
	volume = {543},
	year = {2017}}

@article{goldman2014periodically,
	author = {Goldman, Nathan and Dalibard, Jean},
	journal = {Physical review X},
	number = {3},
	pages = {031027},
	publisher = {APS},
	title = {{Periodically driven quantum systems: Effective Hamiltonians and engineered gauge fields}},
	volume = {4},
	year = {2014}}

@article{bukov2015universal,
	author = {Bukov, Marin and D'Alessio, Luca and Polkovnikov, Anatoli},
	journal = {Advances in Physics},
	number = {2},
	pages = {139--226},
	publisher = {Taylor \& Francis},
	title = {{Universal high-frequency behavior of periodically driven systems: from dynamical stabilization to Floquet engineering}},
	volume = {64},
	year = {2015}}

@article{gauthey1997high,
	author = {Gauthey, FI and Garraway, BM and Knight, PL},
	journal = {Physical Review A},
	number = {4},
	pages = {3093},
	publisher = {APS},
	title = {High harmonic generation and periodic level crossings},
	volume = {56},
	year = {1997}}

@article{de2002high,
	author = {de Morisson Faria, C Figueira and Rotter, I},
	journal = {Physical Review A},
	number = {1},
	pages = {013402},
	publisher = {APS},
	title = {High-order harmonic generation in a driven two-level atom: periodic level crossings and three-step processes},
	volume = {66},
	year = {2002}}

@article{yin2021rabi,
	author = {Yin, Mo-Juan and Wang, Tao and Lu, Xiao-Tong and Li, Ting and Wang, Ye-Bing and Zhang, Xue-Feng and Li, Wei-Dong and Smerzi, Augusto and Chang, Hong},
	journal = {Chinese Physics Letters},
	number = {7},
	pages = {073201},
	publisher = {IOP Publishing},
	title = {{Rabi spectroscopy and sensitivity of a Floquet engineered optical lattice clock}},
	volume = {38},
	year = {2021}}

@article{mishra2021driving,
	author = {Mishra, Utkarsh and Bayat, Abolfazl},
	journal = {Physical Review Letters},
	number = {8},
	pages = {080504},
	publisher = {APS},
	title = {Driving enhanced quantum sensing in partially accessible many-body systems},
	volume = {127},
	year = {2021}}

@article{wang2022sensing,
	author = {Wang, Guoqing and Liu, Yi-Xiang and Schloss, Jennifer M and Alsid, Scott T and Braje, Danielle A and Cappellaro, Paola},
	journal = {Physical Review X},
	number = {2},
	pages = {021061},
	publisher = {APS},
	title = {Sensing of arbitrary-frequency fields using a quantum mixer},
	volume = {12},
	year = {2022}}

@article{wang2021error,
	author = {Wang, Yuan-Sheng and Liu, Bao-Jie and Su, Shi-Lei and Yung, Man-Hong},
	journal = {Physical Review Research},
	number = {3},
	pages = {033010},
	publisher = {APS},
	title = {{Error-resilient Floquet geometric quantum computation}},
	volume = {3},
	year = {2021}}

@article{shi2016quantum,
	author = {Shi, ZC and Wang, W and Yi, XX},
	journal = {Scientific Reports},
	number = {1},
	pages = {22077},
	publisher = {Nature Publishing Group UK London},
	title = {Quantum gates by periodic driving},
	volume = {6},
	year = {2016}}

@book{nielsen2010quantum,
	author = {Nielsen, Michael A and Chuang, Isaac L},
	publisher = {Cambridge university press},
	title = {Quantum computation and quantum information},
	year = {2010}}

@article{rabi1937space,
	author = {Rabi, Isidor Isaac},
	journal = {Physical Review},
	number = {8},
	pages = {652},
	publisher = {APS},
	title = {Space quantization in a gyrating magnetic field},
	volume = {51},
	year = {1937}}

@article{fujii2013introduction,
	author = {Fujii, Kazuyuki},
	journal = {arXiv preprint arXiv:1301.3585},
	title = {{Introduction to the rotating wave approximation (RWA): Two coherent oscillations}},
	year = {2013}}

@article{son2009floquet,
	author = {Son, Sang-Kil and Han, Siyuan and Chu, Shih-I},
	journal = {Physical Review A---Atomic, Molecular, and Optical Physics},
	number = {3},
	pages = {032301},
	publisher = {APS},
	title = {Floquet formulation for the investigation of multiphoton quantum interference in a superconducting qubit driven by a strong ac field},
	volume = {79},
	year = {2009}}

@article{oliver2005mach,
	author = {Oliver, William D and Yu, Yang and Lee, Janice C and Berggren, Karl K and Levitov, Leonid S and Orlando, Terry P},
	journal = {Science},
	number = {5754},
	pages = {1653--1657},
	publisher = {American Association for the Advancement of Science},
	title = {{Mach-Zehnder interferometry in a strongly driven superconducting qubit}},
	volume = {310},
	year = {2005}}

@article{yan2021lamb,
	author = {Yan, Yiying and Ergogo, Tadele T and L{\"u}, Zhiguo and Chen, Lipeng and Luo, JunYan and Zhao, Yang},
	journal = {The Journal of Physical Chemistry Letters},
	number = {40},
	pages = {9919--9925},
	publisher = {ACS Publications},
	title = {{Lamb shift and the vacuum Rabi splitting in a strongly dissipative environment}},
	volume = {12},
	year = {2021}}

@article{silveri2015stuckelberg,
	author = {Silveri, MP and Kumar, KS and Tuorila, J and Li, J and Veps{\"a}l{\"a}inen, A and Thuneberg, EV and Paraoanu, GS},
	journal = {New Journal of Physics},
	number = {4},
	pages = {043058},
	publisher = {IOP Publishing},
	title = {St{\"u}ckelberg interference in a superconducting qubit under periodic latching modulation},
	volume = {17},
	year = {2015}}

@article{satanin2014amplitude,
	author = {Satanin, AM and Denisenko, MV and Gelman, AI and Nori, Franco},
	journal = {Physical Review B},
	number = {10},
	pages = {104516},
	publisher = {APS},
	title = {{Amplitude and phase effects in Josephson qubits driven by a biharmonic electromagnetic field}},
	volume = {90},
	year = {2014}}

@article{blattmann2015qubit,
	author = {Blattmann, Ralf and H{\"a}nggi, Peter and Kohler, Sigmund},
	journal = {Physical Review A},
	number = {4},
	pages = {042109},
	publisher = {APS},
	title = {{Qubit interference at avoided crossings: The role of driving shape and bathcoupling}},
	volume = {91},
	year = {2015}}

@article{ferron2017mesoscopic,
	author = {Ferr{\'o}n, Alejandro and Dom{\'\i}nguez, Daniel and S{\'a}nchez, Mar{\'\i}a Jos{\'e}},
	journal = {Physical Review B},
	number = {4},
	pages = {045412},
	publisher = {APS},
	title = {Mesoscopic fluctuations in biharmonically driven flux qubits},
	volume = {95},
	year = {2017}}

@article{yan2023multiphoton,
	author = {Yan, Yiying and L{\"u}, Zhiguo and Chen, Lipeng and Zheng, Hang},
	journal = {Advanced Quantum Technologies},
	number = {4},
	pages = {2200191},
	publisher = {Wiley Online Library},
	title = {{Multiphoton resonance band and Bloch--Siegert shift in a bichromatically drivenqubit}},
	volume = {6},
	year = {2023}}

@article{forster2015landau,
	author = {Forster, Florian and M{\"u}hlbacher, Max and Blattmann, Ralf and Schuh, Dieterand Wegscheider, Werner and Ludwig, Stefan and Kohler, Sigmund},
	journal = {Physical Review B},
	number = {24},
	pages = {245422},
	publisher = {APS},
	title = {{Landau-Zener interference at bichromatic driving}},
	volume = {92},
	year = {2015}}

@article{shi2021two,
	author = {Shi, Zhi-Cheng and Chen, Ye-Hong and Qin, Wei and Xia, Yan and Yi, XX and Zheng, Shi-Biao and Nori, Franco},
	journal = {Physical Review A},
	number = {5},
	pages = {053101},
	publisher = {APS},
	title = {{Two-level systems with periodic N-step driving fields: Exact dynamics and quantum state manipulations}},
	volume = {104},
	year = {2021}}

@article{deng2016dynamics,
	author = {Deng, Chunqing and Shen, Feiruo and Ashhab, Sahel and Lupascu, Adrian},
	journal = {Physical Review A},
	number = {3},
	pages = {032323},
	publisher = {APS},
	title = {{Dynamics of a two-level system under strong driving: Quantum-gate optimization based on Floquet theory}},
	volume = {94},
	year = {2016}}

@article{xie2010analytical,
	author = {Xie, Qiongtao and Hai, Wenhua},
	journal = {Physical Review A---Atomic, Molecular, and Optical Physics},
	number = {3},
	pages = {032117},
	publisher = {APS},
	title = {Analytical results for a monochromatically driven two-level system},
	volume = {82},
	year = {2010}}

@article{xie2018analytical,
	author = {Xie, Qiongtao},
	journal = {Pramana},
	number = {2},
	pages = {19},
	publisher = {Springer},
	title = {Analytical results for periodically-driven two-level models in relation to Heun functions},
	volume = {91},
	year = {2018}}

@phdthesis{ishkhanyan2019quantum,
	author = {Ishkhanyan, Tigran},
	school = {Universit{\'e} Bourgogne Franche-Comt{\'e}; Institute for Physical Research (Ashtarak)},
	title = {{Quantum two-state level-crossing models in terms of the Heun functions}},
	year = {2019}}

@article{liu2022floquet,
	author = {Liu, Yibo and Mao, Lijun and Zhang, Yunbo},
	journal = {Physical Review A},
	number = {5},
	pages = {053717},
	publisher = {APS},
	title = {{Floquet analysis of extended Rabi models based on high-frequency expansion}},
	volume = {105},
	year = {2022}}

@article{marinho2024approximate,
	author = {Marinho, A and Dodonov, AV},
	journal = {Physica Scripta},
	number = {12},
	pages = {125117},
	publisher = {IOP Publishing},
	title = {{Approximate analytic solution of the dissipative semiclassical Rabi model under parametric multi-tone modulations}},
	volume = {99},
	year = {2024}}

@article{han2024floquet,
	author = {Han, Yingying and Zhang, Shuanghao and Zhang, Meijuan and Guan, Q and Zhang, Wenxian and Li, Weidong},
	journal = {Physical Review A},
	number = {5},
	pages = {053704},
	publisher = {APS},
	title = {{Floquet dynamics of the Rabi model beyond the counter-rotating hybridized rotating-wave method}},
	volume = {109},
	year = {2024}}

@article{chen2022enhanced,
	author = {Chen, Ye-Hong and Miranowicz, Adam and Chen, Xi and Xia, Yan and Nori, Franco},
	journal = {Physical Review Applied},
	number = {6},
	pages = {064059},
	publisher = {APS},
	title = {Enhanced-fidelity ultrafast geometric quantum computation using strong classical drives},
	volume = {18},
	year = {2022}}

@article{yan2017effects,
	author = {Yan, Yiying and L{\"u}, Zhiguo and Luo, JunYan and Zheng, Hang},
	journal = {Physical Review A},
	number = {3},
	pages = {033802},
	publisher = {APS},
	title = {{Effects of counter-rotating couplings of the Rabi model with frequencymodulation}},
	volume = {96},
	year = {2017}}

@article{shirley1965solution,
	author = {Shirley, Jon H},
	journal = {Physical Review},
	number = {4B},
	pages = {B979},
	publisher = {APS},
	title = {{Solution of the Schr{\"o}dinger equation with a Hamiltonian periodic in time}},
	volume = {138},
	year = {1965}}

@article{ho1983semiclassical,
	author = {Ho, Tak-San and Chu, Shih-I and Tietz, James V},
	journal = {Chemical Physics Letters},
	number = {4},
	pages = {464--471},
	publisher = {Elsevier},
	title = {{Semiclassical many-mode Floquet theory}},
	volume = {96},
	year = {1983}}

@article{barone1977floquet,
	author = {Barone, SR and Narcowich, MA and Narcowich, FJ},
	journal = {Physical Review A},
	number = {3},
	pages = {1109},
	publisher = {APS},
	title = {Floquet theory and applications},
	volume = {15},
	year = {1977}}

@article{zeuch2020exact,
	author = {Zeuch, Daniel and Hassler, Fabian and Slim, Jesse J and DiVincenzo, David P},
	journal = {Annals of physics},
	pages = {168327},
	publisher = {Elsevier},
	title = {Exact rotating wave approximation},
	volume = {423},
	year = {2020}}

@article{massa2003new,
	author = {Massa, Enrico and Vignolo, Stefano},
	journal = {extracta mathematicae},
	number = {1},
	pages = {107--118},
	title = {{A new geometrical framework for time-dependent Hamiltonian mechanics}},
	volume = {18},
	year = {2003}}

@article{giscard2023exact,
	author = {Giscard, Pierre-Louis and Foroozandeh, Mohammadali},
	journal = {Computer Physics Communications},
	pages = {108561},
	publisher = {Elsevier},
	title = {Exact solutions for the time-evolution of quantum spin systems under arbitrary waveforms using algebraic graph theory},
	volume = {282},
	year = {2023}}

@article{giscard2020dynamics,
	author = {Giscard, Pierre-Louis and Bonhomme, Christian},
	journal = {Physical Review Research},
	number = {2},
	pages = {023081},
	publisher = {APS},
	title = {{Dynamics of quantum systems driven by time-varying Hamiltonians: Solution for the Bloch-Siegert Hamiltonian and applications to NMR}},
	volume = {2},
	year = {2020}}

@article{giscard2015exact,
	author = {Giscard, P-L and Lui, K and Thwaite, SJ and Jaksch, D},
	journal = {Journal of Mathematical Physics},
	number = {5},
	publisher = {AIP Publishing},
	title = {An exact formulation of the time-ordered exponential using path-sums},
	volume = {56},
	year = {2015}}

@article{kalev2021integral,
	author = {Kalev, Amir and Hen, Itay},
	journal = {New Journal of Physics},
	number = {10},
	pages = {103035},
	publisher = {IOP Publishing},
	title = {{An integral-free representation of the Dyson series using divided differences}},
	volume = {23},
	year = {2021}}

@article{kalev2021quantum,
	author = {Kalev, Amir and Hen, Itay},
	journal = {Quantum},
	pages = {426},
	publisher = {Verein zur F{\"o}rderung des Open Access Publizierens in den Quantenwissenschaften},
	title = {{Quantum algorithm for simulating hamiltonian dynamics with an off-diagonal series expansion}},
	volume = {5},
	year = {2021}}

@article{chen2021quantum,
	author = {Chen, Yi-Hsiang and Kalev, Amir and Hen, Itay},
	journal = {PRX Quantum},
	number = {3},
	pages = {030342},
	publisher = {APS},
	title = {{Quantum algorithm for time-dependent Hamiltonian simulation by permutation expansion}},
	volume = {2},
	year = {2021}}

@article{neto2023basis,
	author = {Neto, Ant{\^o}nio Francisco},
	journal = {Annales de l'Institut Henri Poincar{\'e} D},
	number = {2},
	pages = {383--407},
	title = {{A basis-and integral-free representation of time-dependent perturbation theory via the omega matrix calculus}},
	volume = {11},
	year = {2023}}

@article{hague2016generalized,
	author = {Hague, David A and Buck, John R},
	journal = {IEEE Journal of Oceanic Engineering},
	number = {1},
	pages = {109--123},
	publisher = {IEEE},
	title = {The generalized sinusoidal frequency-modulated waveform for active sonar},
	volume = {42},
	year = {2016}}

@article{hague2020adaptive,
	author = {Hague, David A},
	journal = {IEEE transactions on aerospace and electronic systems},
	number = {2},
	pages = {1274--1287},
	publisher = {IEEE},
	title = {{Adaptive transmit waveform design using multitone sinusoidal frequency modulation}},
	volume = {57},
	year = {2021}}

@article{hague2023characterizing,
  title={Characterizing the narrowband ambiguity function of multi-tone sinusoidal frequency modulated waveforms},
  author={Hague, David A},
  journal={arXiv preprint arXiv:2312.05188},
  year={2023}
}

@article{aravind1984two,
  title={Two-state systems in semiclassical and quantized fields},
  author={Aravind, PK and Hirschfelder, JO},
  journal={The Journal of Physical Chemistry},
  volume={88},
  number={21},
  pages={4788--4801},
  year={1984},
  publisher={ACS Publications}
}

@article{chu1989generalized,
  title={Generalized Floquet Theoretical Approaches to Intense-Field Multiphoton and Nonlinear Optical Processes},
  author={Chu, Shih-I},
  journal={Advances in Chemical Physics: Lasers, Molecules, and Methods},
  volume={73},
  pages={739--799},
  year={1989},
  publisher={Wiley Online Library}
}

@article{lorenzutta1997infinite,
	author = {Lorenzutta, S and Maino, Giuseppe and Dattoli, Giuseppe and Torre, Amalia and Chiccoli, Cesare},
	journal = {Reports on Mathematical Physics},
	number = {2},
	pages = {163--176},
	publisher = {Elsevier},
	title = {{Infinite-variable Bessel functions of the Anger type and the Fourier expansions}},
	volume = {39},
	year = {1997}}

@inproceedings{dattoli1996theory,
	author = {Dattoli, Giuseppe and Torre, Amalia},
	organization = {Aracne Rome},
	title = {{Theory and applications of generalized Bessel functions}},
	year = {1996}}

@article{carson2006notes,
	author = {Carson, John R},
	journal = {Proceedings of the institute of radio engineers},
	number = {1},
	pages = {57--64},
	publisher = {IEEE},
	title = {Notes on the theory of modulation},
	volume = {10},
	year = {2006}}

@article{kuklinski2019identities,
	author = {Kuklinski, Parker and Hague, David A},
	journal = {arXiv preprint arXiv:1908.11683},
	title = {{Identities and properties of multi-dimensional generalized Bessel functions}},
	year = {2019}}

@article{johansson2012qutip,
	author = {Johansson, J Robert and Nation, Paul D and Nori, Franco},
	journal = {Computer physics communications},
	number = {8},
	pages = {1760--1772},
	publisher = {Elsevier},
	title = {{QuTiP: An open-source Python framework for the dynamics of open quantum systems}},
	volume = {183},
	year = {2012}}

@article{wolpert1995estimating,
	author = {Wolpert, David H and Wolf, David R},
	journal = {Physical Review E},
	number = {6},
	pages = {6841},
	publisher = {APS},
	title = {Estimating functions of probability distributions from a finite set of samples},
	volume = {52},
	year = {1995}}

@article{lasserre2001solving,
	author = {Lasserre, Jean B and Zeron, Eduardo S},
	journal = {Applicationes Mathematicae},
	pages = {391--405},
	publisher = {Instytut Matematyczny Polskiej Akademii Nauk},
	title = {{Solving a class of multivariate integration problems via Laplace techniques}},
	volume = {28},
	year = {2001}}

@article{robins2021friendly,
	author = {Robins, Sinai},
	journal = {arXiv preprint arXiv:2104.06407},
	title = {{A friendly introduction to Fourier analysis on polytopes}},
	year = {2021}}

@article{lu2009computable,
	author = {Lu, Yue M and Do, Minh N and Laugesen, Richard S},
	journal = {IEEE Transactions on Signal Processing},
	number = {5},
	pages = {1768--1782},
	publisher = {IEEE},
	title = {{A computable Fourier condition generating alias-free sampling lattices}},
	volume = {57},
	year = {2009}}

@article{dello2019characteristic,
	author = {Dello Schiavo, Lorenzo},
	title = {{Characteristic functionals of Dirichlet measures}},
	year = {2019}}

@article{de2005divided,
	author = {de Boor, Carl},
	journal = {arXiv preprint math/0502036},
	title = {Divided differences},
	year = {2005}}

@article{ullrich1980divided,
	author = {Ullrich, David},
	journal = {Proceedings of the American Mathematical Society},
	number = {1},
	pages = {47--57},
	title = {{Divided differences and systems of nonharmonic Fourier series}},
	volume = {80},
	year = {1980}}

@article{avdonin2001exponential,
	author = {Avdonin, Sergei Anatol'evich and Ivanov, Sergei Alekseevich},
	journal = {arXiv preprint math/0103160},
	title = {{Exponential Riesz bases of subspaces and divided differences}},
	year = {2001}}

@article{bekers2024extending,
	author = {Bekers, Dave J},
	journal = {IEEE Transactions on Aerospace and Electronic Systems},
	publisher = {IEEE},
	title = {{Extending the Multi-Tone Sinusoidal Frequency Modulation Signal Model by Fourier Expansions With Arbitrary Periods}},
	year = {2024}}

@article{giscard2020solutions,
	author = {Giscard, Pierre-Louis},
	title = {{On the solutions of linear Volterra equations of the second kind with sum kernels}},
	year = {2020}}

@inproceedings{felton2023gradient,
	author = {Felton, David G and Hague, David A},
	booktitle = {2023 IEEE Radar Conference (RadarConf23)},
	organization = {IEEE},
	pages = {1--6},
	title = {{Gradient-descent based optimization of constant envelope OFDM waveforms}},
	year = {2023}}

@article{baxterexponential,
	author = {Baxter, BJC and Brummelhuis, Raymond},
	journal = {Foundations of Computational Mathematics},
	title = {{Exponential Brownian motion and divided differences}}}

@article{gomez2023anomalous,
  title={Anomalous Floquet Phases. A resonance phenomena},
  author={G{\'o}mez-Le{\'o}n, {\'A}lvaro},
  journal={arXiv preprint arXiv:2312.06778},
  year={2023}
}

@article{jurkutat2025nuclear,
  title={Nuclear magnetic resonance far off the Larmor frequency: Nonsecular resonances in CaF 2},
  author={Jurkutat, Michael and Safiullin, Kajum and Singh, Pooja and Grage, Stephan L and Haase, J{\"u}rgen and Fine, Boris V and Meier, Benno},
  journal={Physical Review B},
  volume={112},
  number={6},
  pages={L060302},
  year={2025},
  publisher={APS}
}

@article{jeschke2025unexpected,
  title={Unexpected Resonances Could Boost NMR’s Potency},
  author={Jeschke, Gunnar},
  journal={Physics},
  volume={18},
  pages={145},
  year={2025},
  publisher={APS}
}

@article{kropf2012nonsecular,
  title={Nonsecular resonances for the coupling between nuclear spins in solids},
  author={Kropf, Chahan M and Fine, Boris V},
  journal={Physical Review B—Condensed Matter and Materials Physics},
  volume={86},
  number={9},
  pages={094401},
  year={2012},
  publisher={APS}
}

@article{sauer2012entanglement,
  title={Entanglement resonances of driven multi-partite quantum systems},
  author={Sauer, Simeon and Mintert, Florian and Gneiting, Clemens and Buchleitner, Andreas},
  journal={Journal of Physics B: Atomic, Molecular and Optical Physics},
  volume={45},
  number={15},
  pages={154011},
  year={2012},
  publisher={IOP Publishing}
}

@article{giscard2025novel,
  title={Novel frame changes for quantum physics},
  author={Giscard, Pierre-Louis and Faizy, Omid and Bonhomme, Christian},
  journal={arXiv preprint arXiv:2510.04598},
  year={2025}
}

@article{mirri2002modified,
  title={A modified Volterra series approach for nonlinear dynamic systems modeling},
  author={Mirri, Domenico and Luculano, G and Filicori, Fabio and Pasini, Gaetano and Vannini, Giorgio and Gabriella, GP},
  journal={IEEE Transactions on Circuits and Systems I: Fundamental Theory and Applications},
  volume={49},
  number={8},
  pages={1118--1128},
  year={2002},
  publisher={IEEE}
}

@article{oon2024beyond,
  title={Beyond Average Hamiltonian Theory for Quantum Sensing},
  author={Oon, Jner Tzern and Carrasco, Sebastian C and Hart, Connor A and Witt, George and Malinovsky, Vladimir S and Walsworth, Ronald},
  journal={arXiv preprint arXiv:2410.04296},
  year={2024}
}

@article{berns2008amplitude,
  title={Amplitude spectroscopy of a solid-state artificial atom},
  author={Berns, David M and Rudner, Mark S and Valenzuela, Sergio O and Berggren, Karl K and Oliver, William D and Levitov, Leonid S and Orlando, Terry P},
  journal={Nature},
  volume={455},
  number={7209},
  pages={51--57},
  year={2008},
  publisher={Nature Publishing Group UK London}
}

@article{satanin2012amplitude,
  title={Amplitude spectroscopy of two coupled qubits},
  author={Satanin, AM and Denisenko, MV and Ashhab, Sahel and Nori, Franco},
  journal={Physical Review B—Condensed Matter and Materials Physics},
  volume={85},
  number={18},
  pages={184524},
  year={2012},
  publisher={APS}
}

@article{TEDO2025100118,
	author = {S.L. Dongmo Tedo and O.C. Feulefack and J.E. Danga and S.E. {Mkam Tchouobiap} and R.M. {Keumo Tsiaze} and A.J. Fotue and M.N. Hounkonnou and L.C. Fai},
	doi = {https://doi.org/10.1016/j.revip.2025.100118},
	issn = {2405-4283},
	journal = {Reviews in Physics},
	pages = {100118},
	title = {{Harmonic driving and dynamic transitions in the Landau--Zener--St{\"u}ckelberg--Majorana interferometry induced by tunneling flux-driven symmetric transmon qubits}},
	url = {https://www.sciencedirect.com/science/article/pii/S2405428325000176},
	volume = {13},
	year = {2025}}

@book{brunner2017volterra,
  title={Volterra integral equations: an introduction to theory and applications},
  author={Brunner, Hermann},
  volume={30},
  year={2017},
  publisher={Cambridge University Press}
}

\end{document}